\documentclass[review]{elsarticle}

\usepackage{lineno,hyperref,eurosym}

\usepackage{color,xcolor}
\journal{Comm Nonlinear Sci (https://doi.org/10.1016/j.cnsns.2023.107458)}

\begin{document}

\begin{frontmatter}

\title{End-to-end trajectory concept for close exploration \\ of Saturn's Inner Large Moons}

\author[label1]{E. Fantino$^{\star}$}
\author[label1]{B.~M. Burhani}
\author[label1,label3]{R. Flores}
\author[label4]{E.~M. Alessi}
\author[label5]{F. Solano}
\author[label5]{M. Sanjurjo-Rivo}
\address[label1]{Department of Aerospace Engineering, Khalifa University of Science and Technology, P.O. Box 127788, Abu Dhabi (United Arab Emirates)}
\address[label3]{Centre Internacional de M\'etodes Num\`erics en Enginyeria (CIMNE), Gran Capit\`a s/n, 08034 Barcelona (Spain)}
\address[label4]{Istituto di Matematica Applicata e Tecnologie Informatiche "Enrico Magenes", Consiglio Nazionale delle Ricerche, Via Alfonso Corti 12, 20133 Milan (Italy)}
\address[label5]{Department of Aerospace Engineering, Universidad Carlos III de Madrid, 28911 Legan\'es, Madrid (Spain)}
\cortext[]{Corresponding author: elena.fantino@ku.ac.ae (E.~Fantino)}

\begin{abstract}
We present a trajectory concept for a small mission to the four inner large satellites of Saturn. Leveraging the high efficiency of electric propulsion, the concept enables orbit insertion around each of the moons, for arbitrarily long close observation periods. The mission starts with a EVVES interplanetary segment, where a combination of multiple gravity assists and deep space low thrust enables reduced relative arrival velocity at Saturn, followed by an unpowered capture via a sequence of resonant flybys with Titan. The transfers between moons use a low-thrust control law that connects unstable and stable branches of the invariant manifolds of planar Lyapunov orbits from the circular restricted three-body problem of each moon and Saturn. The exploration of the moons relies on homoclinic and heteroclinic connections of the Lyapunov orbits around the L$_1$ and L$_2$ equilibrium points. These science orbits can be extended for arbitrary lengths of time with negligible propellant usage. The strategy enables a comprehensive scientific exploration of the inner large moons, located deep inside the gravitational well of Saturn, which is unfeasible with conventional impulsive maneuvers due to excessive fuel consumption. 
\end{abstract}


\end{frontmatter}

\section{Introduction}
\label{sec:intro}
According to the recently published Decadal Strategy for Planetary Science and Astrobiology 2023-2032 \cite{Decadal2022}, the open scientific questions about our planetary system can only be addressed through the {\it in situ} exploration of the giant planets and their moons. 
In particular, high emphasis has been given to new missions to Saturn, including an Enceladus multiple flyby probe and lander and a Titan orbiter.
Due to the confirmed presence of liquid water, which likely extends beneath the entire surface, Enceladus is considered the best candidate to host life \cite{Decadal2011,Reh2016}: as a matter of fact, in the plumes emanating from the ocean below the icy terrain of the south pole of Enceladus, Cassini detected the presence of chemical elements such as carbon, hydrogen, oxygen and nitrogen, which play a key role in producing amino acids, 
the fundamental constituents of proteins \cite{Porco2006,Spencer2006,Khawaja2019}. The Cassini mission raised new questions regarding Enceladus and the other Inner Large Moons (ILMs) of Saturn, Mimas, Tethys and Dione. As detailed in \cite{Buratti2018,Dougherty2018}, dynamical models of Mimas and Dione suggest the existence of an ocean beneath their surface, but conclusive evidence is still missing. The cause of the red streaks on Tethys is under debate, and so is the alleged young origin of the moons despite their dense cratering. Other scientific open questions regard the relative orbital dynamics, the surface composition and the thermal and internal activity of these bodies.

Numerous follow-up exploration missions of Saturn and its moons have been proposed, for example within the roadmaps of the 2013-2022 National Research Council Decadal Survey \cite{Decadal2011} and ESA Science Programme \cite{CosmicVision2005}. NASA's Dragonfly mission 
to Titan \cite{Lorenz2018} is scheduled for launch in 2026. E$^2$T is a medium-class solar-electric mission to Enceladus and Titan designed in response to ESA's M5 Cosmic Vision Call and aims at a launch opportunity in 2030 \cite{Mitri2018}. The joint NASA-ESA TandEM to Titan and Enceladus \cite{Coustenis2009}, initially planned for a launch in 2020, was eventually cancelled in favor of EJSM-Laplace as the L-class outer Solar System mission candidate. More recent proposals include SILENUS, a multi-lander and orbiter mission to Enceladus \cite{SILENUS2022}, and Moonraker, an Enceladus multiple-flyby mission \cite{Moonraker2022}.

The design of the trajectory to tour the moons of a giant planet can be divided into the following four phases: i) interplanetary transfer; ii) orbit insertion around the planet; iii) transfer and orbit insertion around the first moon of the system; iv) moon tour, including the science orbits around individual bodies and the inter-moon transfers.
All four phases are very demanding in terms of fuel consumption and time of flight. For phases i) and ii), this was demonstrated by Cassini/Huygens, the only Saturn orbiter to date. Propellant accounted for more than 50\% of the launch mass of the probe (5655 kg). Thanks to a characteristic launch  energy of 18 km$^2$/s$^2$ and a sequence of planetary gravity assists (GAs) with Earth, Venus and Jupiter, the spacecraft (S/C) reached Saturn in 6.7 years with a consumption of 1100 kg of fuel. The hyperbolic excess speed was 5.55 km/s. The insertion into an elliptical orbit ($2 \cdot 10^4$ km $\times$ $9.3 \cdot 10^6$ km) around the planet was performed by means of a braking maneuver at the pericenter of the arrival hyperbola, which produced a velocity variation $\Delta V$ of 622 m/s. The capture and the subsequent pericenter raising maneuver consumed 1150 kg of propellant, with the remainder supporting the 13-year planetary tour consisting of almost 300 orbits around Saturn with multiple flybys of Titan and other major moons \cite{Peralta1995}. In summary, a large amount of propellant was burnt to decelerate the S/C so that it could be captured by the planet's gravity upon arrival from interplanetary space. 
Inserting a S/C into orbit around one or more moons deeply immersed in the gravity well of the planet, i.e., phases iii) and iv) above, is even more challenging from a dynamical point of view. This endeavour has never been attempted. However, answering the scientific questions listed previously requires extended observations of these bodies, hence the need for probes orbiting them. 

The first concept study of a moon tour has been the Petit Grand Tour (PGT) of the Galilean moons \citep{Koon2000,Koon2002}, where dynamical systems methods are employed to design transfers between libration point orbits (LPOs)
of the  Circular Restricted Three-Body Problems (CR3BPs) formed by Jupiter and individual moons. Coupling distinct CR3BPs allows to patch trajectories belonging to the two systems. The connections are sought between unstable and stable hyperbolic invariant manifolds (HIMs) of periodic orbits around the collinear libration points 
\citep{Gomez2001b,Gomez2003}. A tour of a system of  moons can be designed with the resonance hopping technique \cite{Anderson2005,Strange2009,Palma2016}, which allows to patch consecutive orbits in mean motion resonance with a given moon. The periodic perturbation caused by the moon can be exploited to change the features of the next encounter or to redirect the 
S/C towards another target. This approach can be enhanced by applying small maneuvers in proximity of the GAs \cite{Campagnola2010c,Campagnola2010a}. The resonance hopping concept can be envisaged also within the CR3BP \cite{Lantoine2009,Campagnola2010b,Campagnola2014}: by means of the Tisserand-Poincar\'e graph, it is possible to link different CR3BPs and identify families of periodic orbits leading to a resonant hopping tour. The case of Saturn was analysed by \cite{Brown2008,Russell2009,Strange2009,Palma2016}. The use LPOs in this context was proposed by \cite{Brown2008} based on the observational proof of the existence of Trojans in the Saturn-Dione and Saturn-Tethys systems. Lantoine and Russell \cite{Lantoine2011} designed a resonance hopping tour in the Jovian system by connecting resonant orbits through HIMs of halo orbits of planet-moon CR3BPs.
A different approach searches for stable orbits around the moon. The challenge is to model and properly exploit the gravitational field of the object to identify suitable frozen orbits. Along this line, Russell and Brinckerhof \cite{Russell2009b} proposed mid-inclination, high-altitude, stable, eccentric frozen orbits, obtained by doubly-averaged equations of motion of the third-body effect. They considered the Jovian moons as a test case, but a similar analysis can be applied to the Saturn system, as shown by \cite{Lara2010}. 

The main objective of this contribution is to demonstrate a complete trajectory concept from Earth departure to a tour of the ILMs with minimum fuel consumption. 
The design methodology takes into account the limitations of power and propulsion technologies in a Saturn mission, the cost of launching into deep space and the expected scientific return of a tour of the major moons of Saturn. The study makes use of state-of-the-art methodologies, including patched conics, GAs, optimal control and dynamical systems tools. The aim is to minimize cost ($\Delta V$) and mass while ensuring a mission timeline of an acceptable duration. The propulsion system selected is electrical,
and the power is supplied by radioisotope thermoelectric generators. As a result, the onboard resources are extremely scarce, which limits the available thrust. For this reason, 
the gravitational field of planets and moons is utilized to a large extent so as to minimize the use of the thruster and limit the time of flight. The transfer strategy is carefully planned and leverages the natural dynamics of the involved systems: the interplanetary trajectory and the orbit insertion at Saturn make use of GAs, whereas the tour of the ILMs is based on low-energy (LE) orbits associated with the L$_1$ and L$_2$ equilibrium points of the CR3BPs with Saturn and individual moons as primaries.

The main element of originality of the work resides in the unprecedented concept of achieving orbit around the four moons of Saturn, using only low-thrust (LT) propulsion and gravitational assistance. Preliminary, partial versions of this work can be found in \cite{Fantino2019,Fantino2021,Burhani2021}. Here, we present a complete plan including re-designed interplanetary trajectory with global optimization techniques and revised Saturn Orbit Insertion (SOI) and transfer to Dione, the first moon of the tour. In addition, we study the observational performance of the science orbits around the four targets. 

The paper is organised as follows: Sect.~\ref{sec:model} defines the dynamical models adopted in the study, Sect.~\ref{sec:interplanetary} illustrates the design of the interplanetary trajectory, Sect.~\ref{sec:insertion} focuses on the gravity-assisted orbit insertion at Saturn, while Sect.~\ref{sec:tour} is devoted to the transfer to the ILMs and the design of the tour of Dione, Tethys, Enceladus and Mimas. A discussion of the results can be found in 
Sect.~\ref{sec:disc}, while the strengths and limitations of the investigation as well as an anticipation of future work are the topics of Sect.~\ref{sec:conclu}. The following abbreviations are often used in tables and figures: S = Saturn, Ti = Titan, Di = Dione, Te = Tethys, En = Enceladus, Mi = Mimas.

\section{Dynamical models}
\label{sec:model}

\subsection{Interplanetary trajectory}
The interplanetary portion of the trajectory is a gravity-assisted itinerary alternating ballistic and LT arcs between planetary encounters. The dynamical model is the two-body problem (2BP) with the patched conics approximation. The GAs are with Venus and Earth. They are unpowered and designed assuming zero radius for the planetary spheres of influence (ZRSI); hence, the pericenter altitude of the flyby hyperbolas is a free parameter.
Planetary orbits are assumed coplanar, and the trajectory of the S/C is also 2D. The positions and velocities of the planets are obtained from JPL NAIF-SPICE kernels \cite{NAIF-SPICE} in the Sun-centered International Celestial Reference Frame (ICRF) and are projected on the $xy$-plane as follows:
\begin{itemize} 
\item the projection of the position vector ${\bf r}$ = ($r_x$, $r_y$, $r_z$) preserves the heliocentric distance, whereas the polar angle from the $x$ axis is computed as the 2-argument arctangent of $r_y$ and $r_x$; 
\item the magnitude of the 2D-projected velocity vector is equal to the magnitude of ${\bf v}$ =  ($v_x$, $v_y$, $v_z$), and the flight path angle $\gamma$ on the ecliptic plane is equal to the corresponding quantity in 3D: $\tan \gamma =  \mid {\bf r} \times {\bf v}\mid / ( {\bf r} \cdot {\bf v})$.
\end{itemize}  
For the propelled phases of the transfer, a constant thrust magnitude $T$ is assumed. 
Hence, the dynamical equations for the S/C in the presence of thrust are:
\begin{equation}
\left\{ \begin{array}{lcl}
\dot{\bf r} & = & {\bf v}\\
\dot{\bf v} & = & \displaystyle -\frac{GM}{r^3}{\bf r} + \frac{T}{m}{\bf u}\\
\dot{m} & = & \displaystyle -\frac{T}{g_0I_{sp}}\\
\end{array} \right.
\label{eq:eqs_2BP_Thrust}
\end{equation}
$GM$ being the gravitational parameter of the Sun, $m$ the mass of the S/C, $I_{sp}$ the specific impulse of the thruster (assumed constant), ${\bf u}$ the instantaneous direction of thrust and $g_0$ = 9.81 m/s$^2$. 
The relevant physical data for the Sun, Venus, Earth and Saturn are listed in Table~\ref{tab:SS}. A minimum altitude of 200 km is imposed for the GAs with Venus and Earth.
\begin{table}[h!]
\caption{Relevant physical data for the Sun and the three planets involved in the study: gravitational parameter $GM$ and mean radius $R$.}
\label{tab:SS}
\centering
\begin{tabular}{lrr}
\hline
Body		 		& $GM$          								&  $R$ 	  \\
						& ($10^5$ km$^3$/s$^2$)   & (km)    \\ \hline
Sun      		& $13271$                	&  -  \\    
Venus     	& $3.2486$      								&  6051.8   \\   
Earth     	& $3.9850$      								&  6371.0    \\     
Saturn	    & $379.22$    								&  60268  \\ \hline 
\end{tabular}
\end{table} 

\subsection{Saturn orbit insertion }
The encounter of the S/C with Saturn occurs at the point of intersection between the respective heliocentric trajectories. The hyperbolic excess velocity ${\bf v}_{\infty}$ of the S/C relative to Saturn is computed as the difference between the velocities of the S/C and Saturn. In this way, the pericenter radius of the planetocentric hyperbola is a free parameter.
The orbit insertion and the descent towards the ILMs are carried out through a sequence of unpowered GAs with Titan followed by a propelled arc. The GAs are modelled with the patched conics approximation assuming ZRSI for this moon. As a result, the pericenter altitude of the Titan-centered hyperbolas is a free parameter.

\subsection{Descent towards the ILMs and moon tour}
The descent from Titan to the ILMs is a propelled transfer to the vicinity of Dione with suitable conditions to start the tour. Both the transfer to Dione and the tour are designed in 2D in the equatorial plane of Saturn. The tour alternates inter-moon propelled transfers and science orbits computed in the four Saturn-moon CR3BPs and linked at suitably-defined Poincar\'e sections where the three-body states of the S/C are used to determine osculating 2BP orbits relative to Saturn (see also \cite{Fantino2017}). 
In the descent to Dione and in the inter-moon transfers, the motion of the S/C is determined by Eq.~\ref{eq:eqs_2BP_Thrust} with $GM$ equal to the gravitational parameter of Saturn 
 and ${\bf r}$, ${\bf v}$ representing the state of the S/C in the Saturn-centered inertial reference frame.
In the science phases, the S/C is subject to the gravitational attraction of Saturn (mass $m_1$) and a moon (mass $m_2$), and the two primaries move in circular orbits around their centre of mass. The equations of motion of the S/C are written with respect to the barycentric synodic reference frame of the primaries, with these two bodies on the $x$ axis \cite{Szebehely1967}. Using the gravitational constant $G$, the sum $m_1+m_2$ of the masses and the distance $d$ between the primaries as reference magnitudes, the mean motion of the orbits of the primaries takes unitary value and their orbital period equals $2\pi$. The larger (Saturn) and the smaller (moon) bodies are at ($\mu$,0,0) and ($\mu$-1,0,0), respectively, $\mu$  being the mass ratio $m_2/(m_1+m_2)$ of the system. Table~\ref{tab:param} lists the main physical and dynamical parameters of the ILMs and Titan and the relevant data of the four CR3BPs. 

\begin{table}[h!]
\caption{Main physical and orbital parameters of the four ILMs and Titan (mean physical radii $R$, orbital periods $P$, orbital radii $d$), and relevant data of the four Saturn-moon CR3BPs (mass ratios $\mu$ and $x$-coordinates of L$_1$ and L$_2$ in the moon-centered synodic reference frames).}
\label{tab:param}
\centering
\begin{tabular}{lrrrrrr}
\hline
Moon		& $R$     &  $P$ 			& $d$ 				& $\mu$        & $x$(L$_1$)  		& $x$(L$_2$)     \\
		    & (km)    & (day)     & (km)        & ($10^{-6})$       & (km) 				  &  (km)		  \\ \hline
Mi     & 198      &  0.9424    & 186000    	& 0.06599      &  521  & -521  \\    
En     & 252      &  1.370    & 238000   		& 0.18993      &  947  & -950  \\   
Te     & 533      &  1.888    & 295000    	& 1.08660      &  2097  & -2107  \\     
Di	    & 561     &  2.737    & 377000  		& 1.92799      &  3245  & -3262  \\ 
Ti      & 2575 & 15.95 & 1222000 & - & -& - \\ \hline 
\end{tabular}
\end{table} 


\section{Design and optimization of the interplanetary trajectory}
\label{sec:interplanetary}
To enable practical orbit insertion around the inner moon of Saturn using only low thrust, our mission concept relies on an interplanetary trajectory with a low hyperbolic excess speed at arrival. Fantino et al. \cite{Fantino2020} showed the possibility of having an interplanetary trajectory with an arrival hyperbolic excess speed of 1 km/s. As explained in Sect.~\ref{fig:tour}, this low relative velocity simplifies unpowered capture via a Titan flyby. In \cite{Fantino2020}, the mission design was based on LT electric propulsion and a single unpowered GA. The parameters of the propulsion system corresponded to the NEXT ion engine \cite{VanNoord2007} with 25 mN of thrust, 1400 s of specific impulse and 600 W of power consumption. The initial mass of the S/C was 1500 kg with a characteristic launch energy of 67.25 km$^2$/s$^2$. The two legs of the itinerary, i.e., Earth-Jupiter and Jupiter-Saturn, were optimized separately and without considering phasing constraints for the planets. In the first leg, the thruster was continuously fired in the direction of motion. In the second leg, a locally-optimal guidance law was used to determine the direction of thrust that allows to reach Saturn with a hyperbolic excess speed of 1 km/s. The transfer duration to Saturn was 13 years, consuming 367 kg of propellant (propellant mass fraction = 0.367).

In the present work, the interplanetary trajectory design is refined to obtain a more realistic solution. The launch characteristic energy is reduced and  phasing constraints for the GAs and arrival at Saturn are included, considering the ephemeris of the planets. Additionally, to accommodate the propellant required for transfers inside the Saturn system, the launch mass is increased to 1500 kg, and the parameters of the propulsion system are changed to those of a PPS X00 Hall thruster \cite{Vaudolon2018}. This propulsion system generates acceptable thrust levels (up to 60 mN) with limited power consumption (below 1000 W). Three points from the performance envelope, with input power comparable to the value assumed in \cite{Fantino2020}, have been considered for the trajectory analysis (Table~\ref{tab:hall}). 
\begin{table}[t!]
\caption{Thrust $T$, input power $P_{in}$ and specific impulse $I_{sp}$ of three settings of the PPS X00  thruster studied \cite{Vaudolon2018}.}
\label{tab:hall}
\centering
\begin{tabular}{cccc}
\hline
Setting & $T$		& $P_{in}$     				&  $I_{sp}$     \\
 & (mN)		  & (W)             & (s) \\ \hline
1 & 36    &  640                & 1600 \\
2 & 40    & 650                 & 1450 \\
3 & 50    & 850                 & 1400 \\ \hline    
\end{tabular}
\end{table} 

The optimization of the trajectory is performed using an automated hybrid optimal control method.  This is a two-stage approach \cite{Morante2019}: 1) in the first step, a multi-objective heuristic algorithm explores the global solution space using a surrogate dynamical model, which assumes a predefined shape for the thrust arcs (generalized logarithmic spirals \cite{Roa2016b}). The first step determines the integer optimization variables (i.e., the number and sequence of GAs); 2) in the second step, a single-objective direct collocation method includes a full dynamical model and complex constraints for obtaining a higher-fidelity solution; a result from the previous step is used as an initial guess.

The heuristic global search (step 1) has been conducted assuming a maximum thrust level of 50 mN and a specific impulse of 1400 s (setting \#3 in Table \ref{tab:hall}). A maximum interplanetary transfer duration of 15 years is assumed, together with a minimum characteristic launch energy of 16 km$^2$/s$^2$. The minimum characteristic launch energy that met the 15-year constraint was 27 km$^2$/s$^2$, which is the value used for all the results presented henceforth. Itineraries involving intermediate GAs with Earth, Venus and Jupiter were considered. However, because the sequences including Jupiter required excessive transfer times, they are not presented here. A GA with Jupiter is usually desirable to reduce transfer times. It becomes much less attractive when the 1 km/s relative velocity constraint on arrival is introduced. A strong boost from Jupiter tends to increase the arrival excess speed, requiring in turn more propellant and time for the interplanetary transfer (i.e., the propulsion system needs more time to bleed mechanical energy from the spacecraft before arrival).

\begin{figure}
\centering
\includegraphics[width=0.7\textwidth]{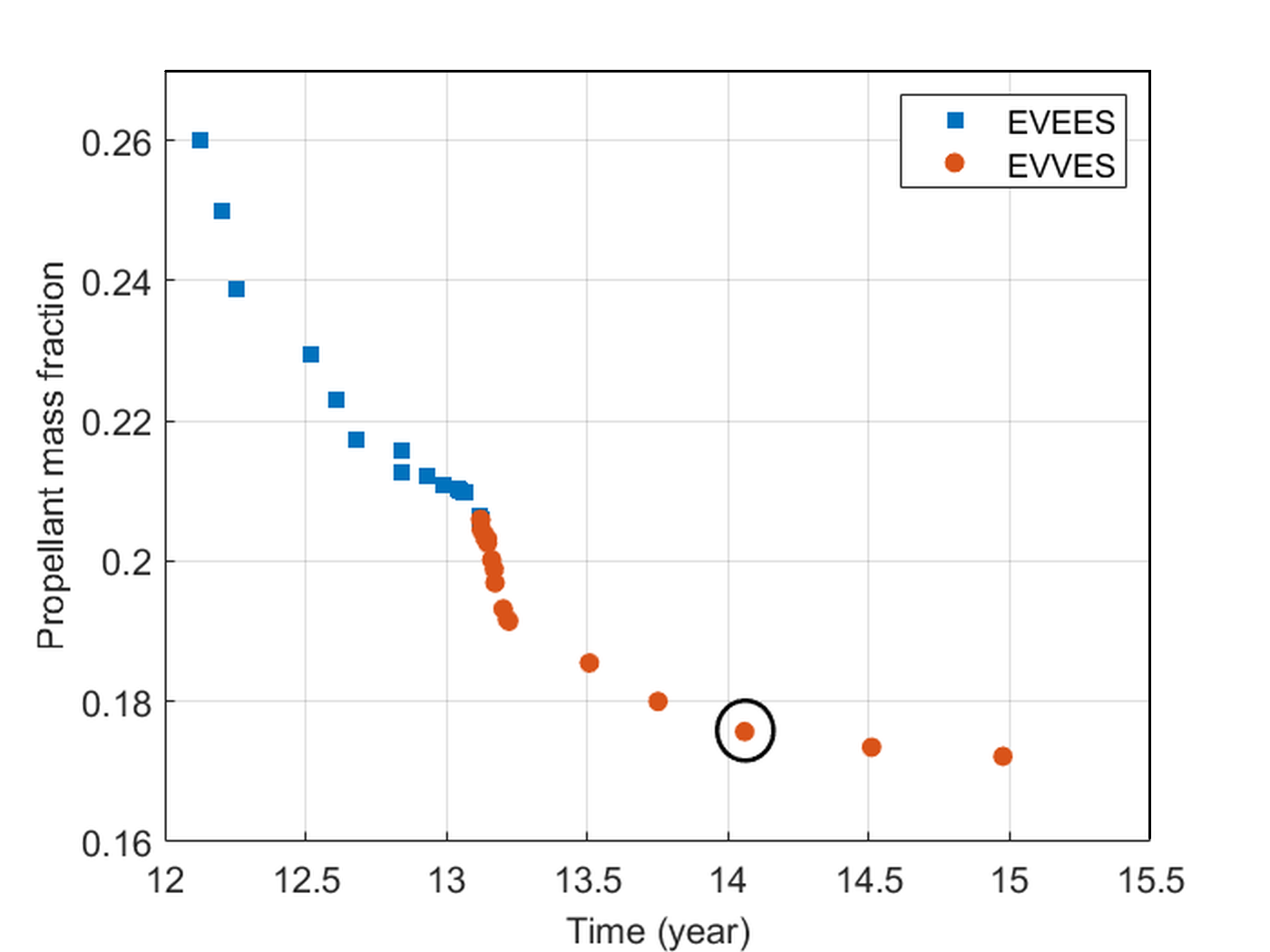} 
\caption{Propellant mass fraction vs. total transfer time for the optimal solutions of the global heuristic search.}
\label{fig:pareto}
\end{figure}

The heuristic search is performed using a multi-objective genetic algorithm (NSGA-II \cite{NSGAII}) to minimize the total transfer time and the propellant mass fraction. The algorithm has been run with populations of 100 individuals, which evolve over 500 generations. The obtained optimal trajectories correspond to two itineraries: Earth-Venus-Earth-Earth-Saturn (EVEES) and Earth-Venus-Venus-Earth-Saturn (EVVES). They are depicted in the Pareto front of Fig.~\ref{fig:pareto}. The transfer times are substantially longer (above 12 years) than for conventional mission concepts. This is a direct consequence of the requirement for a small relative arrival velocity. It results in a slow rate of approach to Saturn, automatically making the last interplanetary segment very long.

\begin{figure}[h!]
\centering
\includegraphics[width=0.485\textwidth]{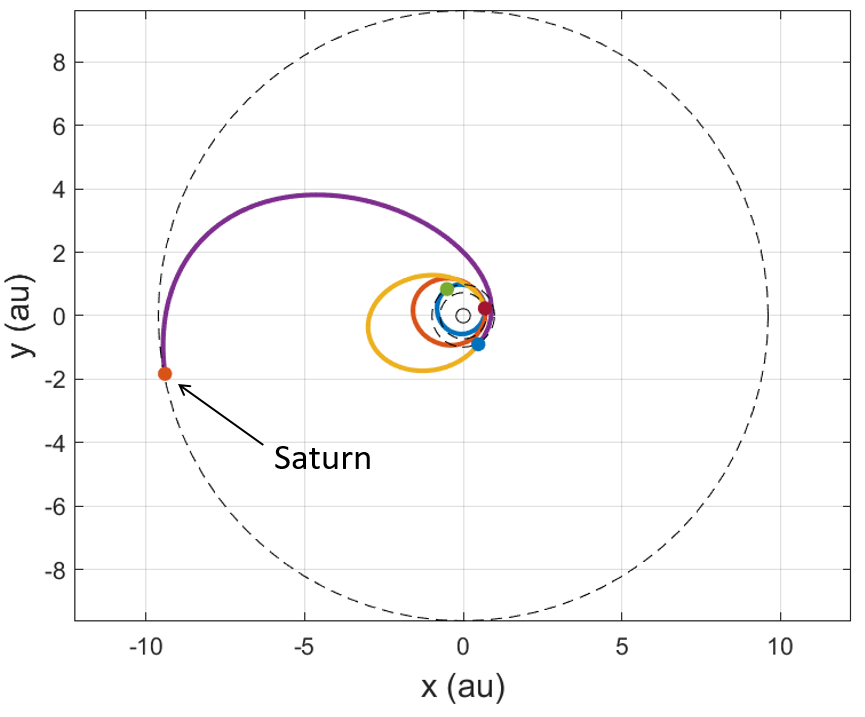} \includegraphics[width=0.497\textwidth]{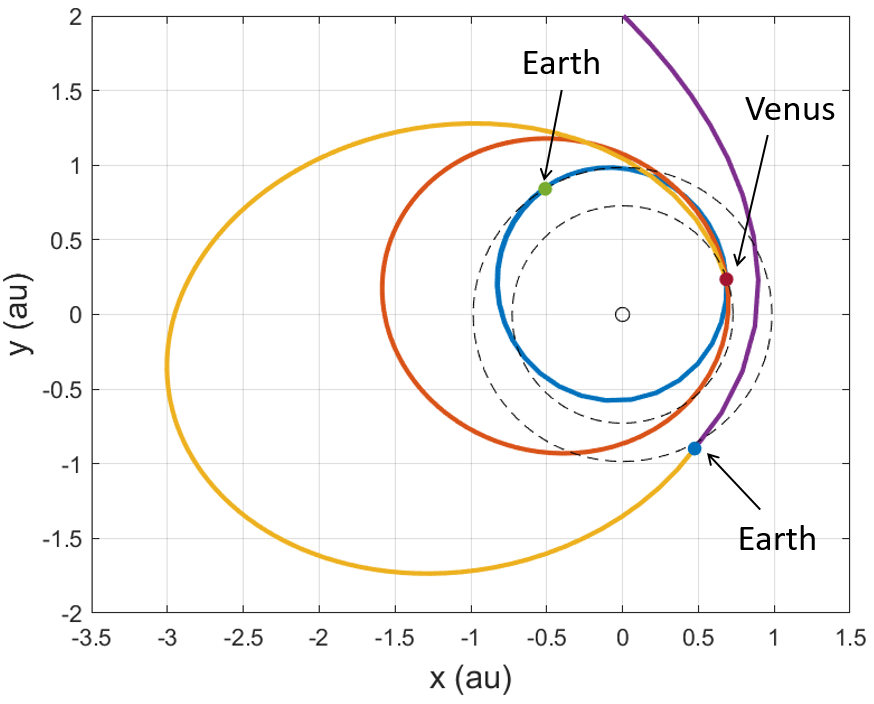} 
\caption{The optimal EVVES trajectory (left) and expanded view of its inner portion (right).}
\label{fig:traj}
\end{figure}

\begin{figure}[h!]
\centering
\includegraphics[width=0.48\textwidth]{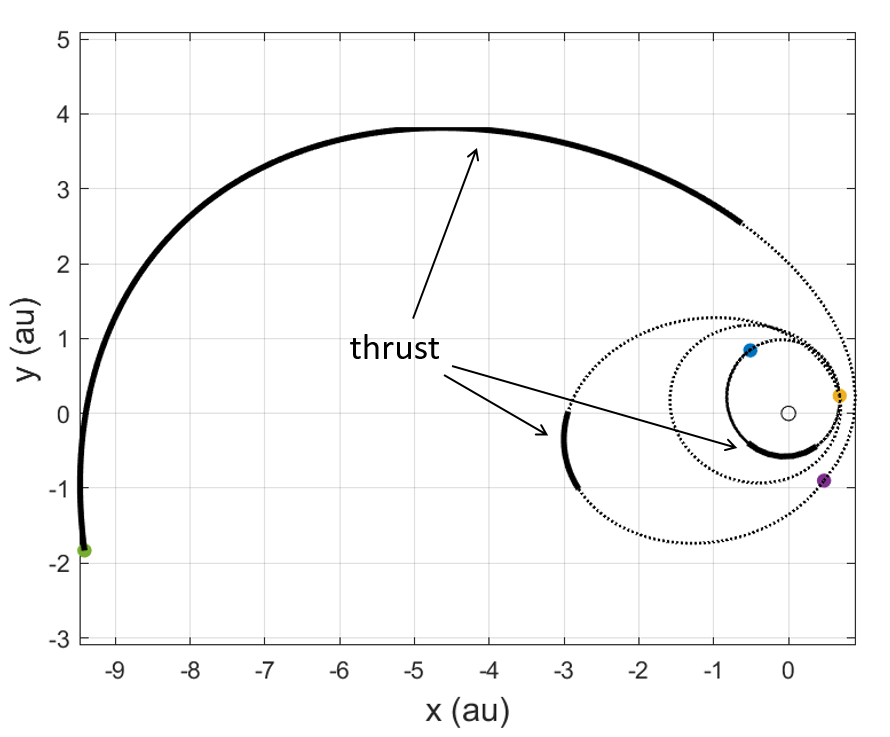} 
\caption{Propelled (thick lines) and ballistic (dotted lines) arcs of the optimal trajectory.}
\label{fig:traj_thrust}
\end{figure}

\begin{figure}[h!]
\centering
\includegraphics[width=0.49\textwidth]{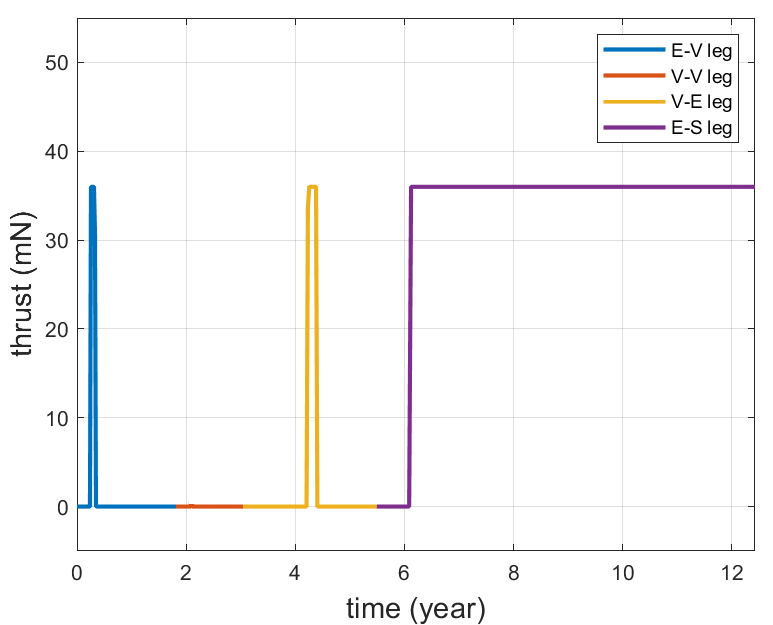} \includegraphics[width=0.49\textwidth]{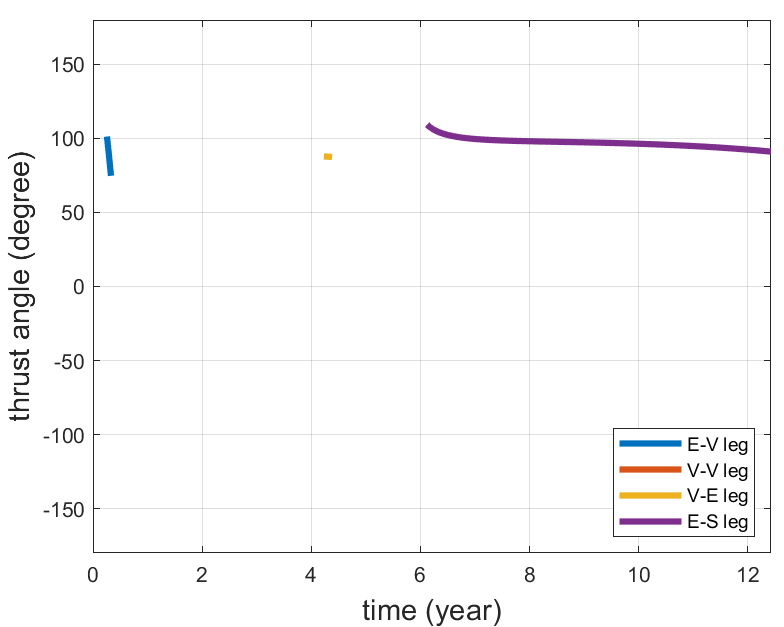} 
\caption{Thrust magnitude (left) and thrust angle (right) histories for the optimal trajectory.}
\label{fig:thrust_angle_mag}
\end{figure}

The EVVES solutions are characterized by lower propellant consumption and longer transfer times than the EVEES trajectories. 
When used as initial guesses for the second optimization stage, EVEES trajectories yield converged solutions only for the highest thrust setting in Table \ref{tab:hall} (50 mN). EVVES trajectories, on the other hand, can be used for all settings (36 to 50 mN). The solution circled in black in Fig.~\ref{fig:pareto}, corresponding to a transfer time of 14.06 years and a propellant mass fraction of 0.176, strikes a good compromise between the two objectives and has been further optimized with the direct method (step 2). Applying different relative weights ($w$) to the transfer time $t_f$ and the final mass $m_f$ in the objective function $J$
\begin{equation}
J = -m_f + w \cdot t_f,
\label{eq:obj_function}
\end{equation}
and decreasing the maximum thrust yields the results reported in Table~\ref{tab:optimal}: for each of the seven solutions, the table lists the throttle setting, transfer time $t_f$,  total impulse $I = \int_0 ^{t_T} |{\bf a}_T|$ {d}$\tau$ (with ${\bf a}_T$ denoting the thrust acceleration), duration $t_T$ of the thrust interval, propellant mass fraction $pmf$, launch date $T_L$ and arrival date $T_A$. Solution \#4 (highlighted in boldface) is the best compromise between propellant consumption and transfer time. Figure~\ref{fig:traj} illustrates the complete trajectory and an expanded view of its inner portion, while Fig.~\ref{fig:traj_thrust} highlights the propelled arcs (thick solid lines) and ballistic segments (dotted lines). The thrust magnitude and thrust angle histories are plotted in Fig.~\ref{fig:thrust_angle_mag}, where an angle of zero corresponds to thrusting in the  circumferential direction.

\begin{table}[h!]
\caption{Optimal solutions obtained from the selected trajectory in Fig.~\ref{fig:pareto} through the direct collocation method for different maximum thrust levels and cost function weight parameter $w$}.
\label{tab:optimal}
\centering
{\footnotesize
\begin{tabular}{rrrrrrrrr}
\hline
Sol. & Thrust & $w$ & $t_f$ & $I$ & $t_T$ & $pmf$ & $T_L$ & $T_A$ \\
	\#	      & setting & (kg/year) & (year) & (km/s) & (year)       & & (yyyy/mm/dd) & (yyyy/mm/dd) \\ \hline
1 &   1  &   0    & 14.8           & 5.83         & 6.50     & 0.311   & 2023/02/09 & 2037/12/22 \\								
2 &   1  & 188    & 11.8           & 7.59         & 8.10     & 0.383   & 2023/02/02 & 2034/11/20 \\								
3 &   1  & 157    & 12.0           & 7.86         & 8.34     & 0.394   & 2025/01/01 & 2036/12/25 \\								
{\bf 4} & {\bf  1}  &  {\bf 118} & {\bf 12.3} & {\bf 6.16}   & {\bf 6.81} & {\bf 0.324} & {\bf 2028/01/21} & {\bf 2040/05/24} \\		5 &   2  & 0    & 14.8           & 6.18         & 6.02     & 0.352   & 2023/01/01 & 2037/10/23 \\								
6 &   2  & 188  & 12.0           & 6.57         & 6.37     & 0.370   & 2028/01/20 & 2040/01/02 \\								
7 &   3  & 0    & 14.2           & 5.60         & 4.43     & 0.335   & 2023/02/16 & 2037/04/27 \\ \hline    
\end{tabular}}
\end{table} 


\section{Saturn orbit insertion}
\label{sec:insertion}
The low hyperbolic excess speed $v_{\infty}$ of 1 km/s enables unpowered capture via a GA with Titan. This is important, because the concept of the mission (low-thrust propulsion only) does not allow for impulsive maneuvers. While it is possible to achieve ballistic SOI with substantially higher arrival velocities, the very long period of the post-GA orbit makes this option impractical \cite{Fantino2020}. On the other hand, lower excess speeds offer little advantage for orbit insertion ---the Saturn inbound hyperbola is very similar to a parabola--- but result in longer interplanetary transfer times with increased propellant consumption. A relative velocity of 1 km/s offers a reasonable trade-off, and has been retained for all the calculations. Obviously, it would be possible to include $v_{\infty}$ as a free optimization parameter in order to improve the results. However, to demonstrate the mission concept and keep the discussion as simple as possible, this refinement was deemed unnecessary.

The GA with Titan is performed in the inbound leg of the Saturn-centered arrival hyperbola. Assigning a pericenter  altitude of 1295 km  (the lowest height reached by Cassini above this moon was 1000 km) results in an elliptical orbit with a period of approximately 80 days or 5 orbits of Titan, allowing the S/C to perform another GA with this moon. A pericenter altitude of 3775 km in this second encounter leads to a resonant orbit with twice the period of Titan. If the third close encounter with the moon occurs at an altitude of 2258 km, the post-GA orbit is in a 1:1 resonance state with Titan. A passage 2690 km above the surface in the fourth GA places the pericenter of the post-GA orbit at the radius of Dione.
\begin{figure}[h]
\centering
\includegraphics[width=0.6\textwidth]{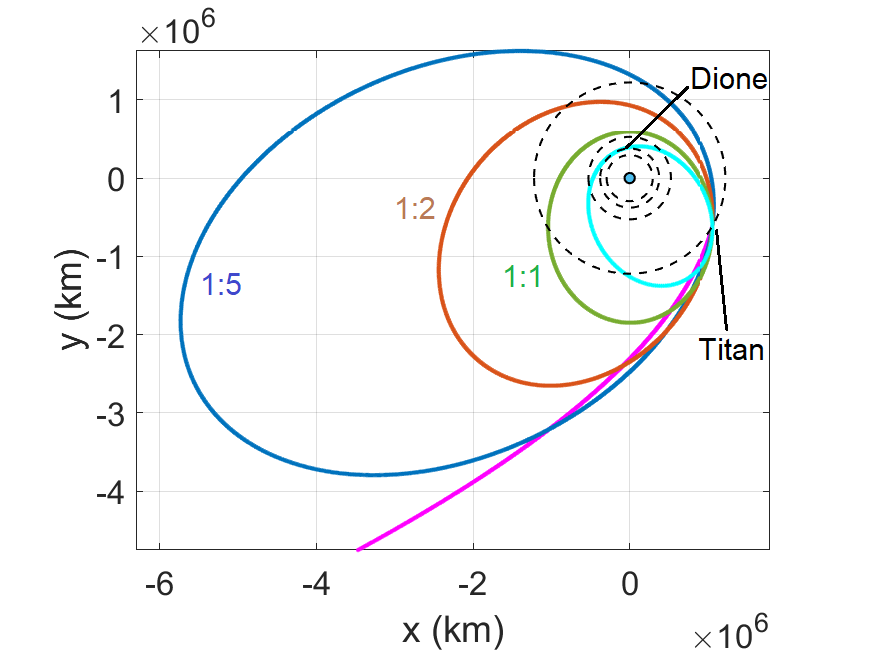} 
\caption{Gravitational capture at Saturn and resonant GAs with Titan: arrival hyperbola (magenta) and post-GA orbits (5:1 in blue, 2:1 in orange, 1:1 in green, orbit with pericenter at the radius of Dione in cyan).}
\label{fig:Titan_GAs}
\end{figure}
Figure~\ref{fig:Titan_GAs} depicts the transfer from the initial hyperbola through the four GAs to the final orbit. The notation $m$:$n$ means that $m$ orbital periods of the S/C equal $n$ orbital periods of Titan. The pericenter altitude $h_{\pi}$ of the four GAs and the characteristics of the post-GA orbits (semimajor axes $a^+$, eccentricities $e^+$ and periods $P^+$) are reported in Table~\ref{tab:flybys}. Note that the resonant state causes the four Titan-centered hyperbolas to have the same  $v_{\infty}$ of 2.962 km/s. 
\begin{table}[t!]
\caption{Pericenter altitudes $h_{\pi}$ of the four GAs with Titan and semimajor axes $a^+$, eccentricities $e^+$ and periods $P^+$ of the post-GA orbits.}
\label{tab:flybys}
\centering
\begin{tabular}{crrrr}
\hline
GA\#		& $h_{\pi}$     				&  $a^+$    	& $e^+$  		& $P^+$     \\
		    & (km)                  & ($10^6$ km)  					& -  				   & (day) \\ \hline
1	    & $1295$   				       &  3.5739     &  0.7195  &  79.79  \\ 
2     & $3775$    			        &  1.9399     &  0.5602  & 31.91  \\     
3     & $2258$    			        &  1.2220     &  0.5126  & 15.96  \\   
4     & $2690$    			        &  0.9298     &  0.5943  & 10.59  \\ \hline    
\end{tabular}
\end{table} 

\section{Descent to Dione and moon tour}
\label{sec:tour}
\begin{figure}[h]
\centering
\includegraphics[width=0.9\textwidth]{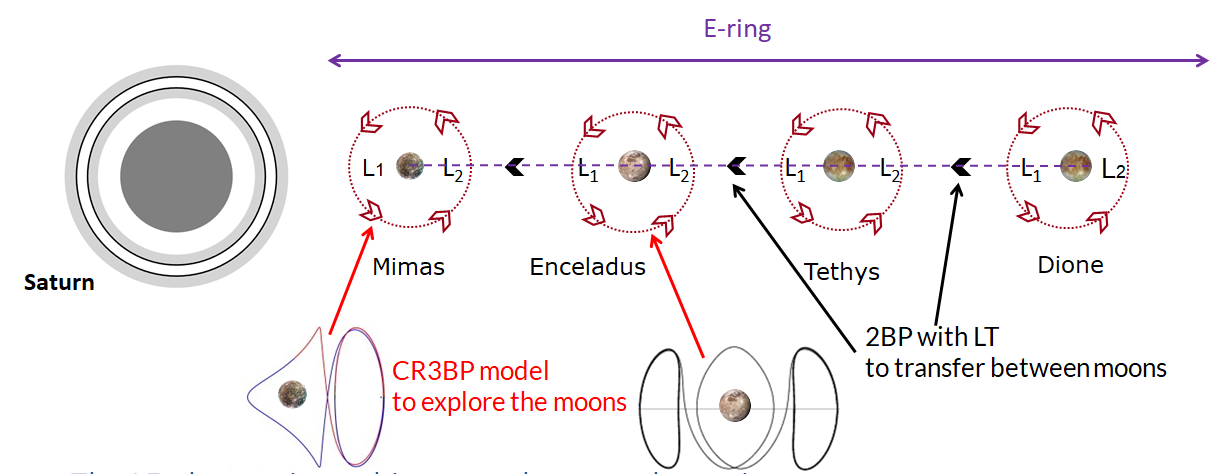} 
\caption{The moon tour: LE transfers as science orbits in the vicinity of the moons and LT Saturn-centered trajectories to move between moons.}
\label{fig:tour}
\end{figure}

The tour alternates LE orbits in the vicinity of the moons (Sect.~\ref{sec:science}) and LT trajectories in the inter-moon transfers (Sect.~\ref{sec:inter}), proceeding from the outermost (Dione) to the innermost (Mimas) body. The concept of the tour is sketched in Fig.~\ref{fig:tour}. The descent from Titan to Dione has been designed with the same technique as the inter-moon transfers. For this reason, it is described in Sect.~\ref{sec:inter}. The descent to Dione after the resonant flybys over Titan could be shortened using GAs with Rhea. This possibility has not been considered for the sake of brevity and simplicity. Furthermore, due to the small mass of Rhea, the effect is limited compared with the Titan GAs. For a proof of concept, neglecting the influence of Rhea gives a conservative estimate of mission length and propellant budget, so it is acceptable. It should be included, however, in a detailed mission design.

\subsection{Science orbits}
\label{sec:science}
\begin{figure}[h!]
\centering
\begin{tabular}{cc}
\includegraphics[width=0.49\textwidth]{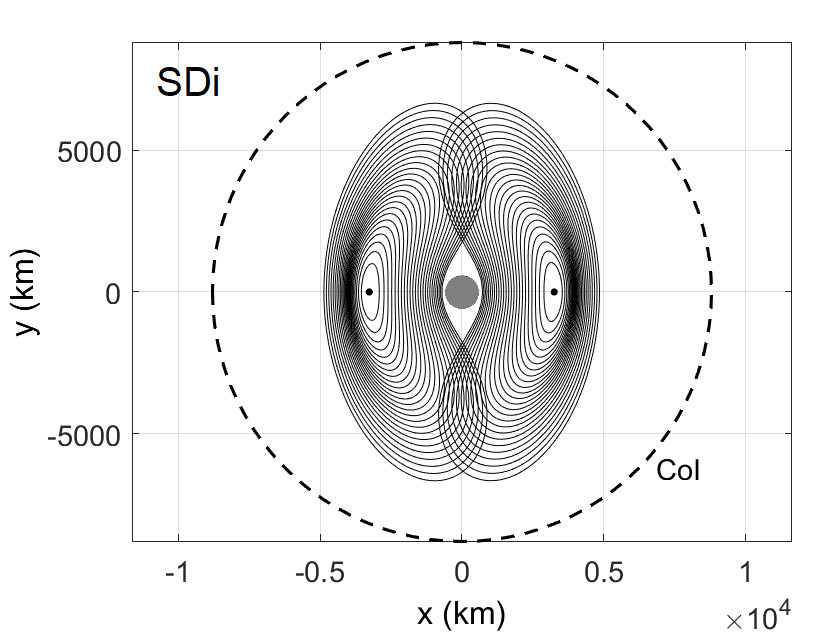} & \includegraphics[width=0.49\textwidth]{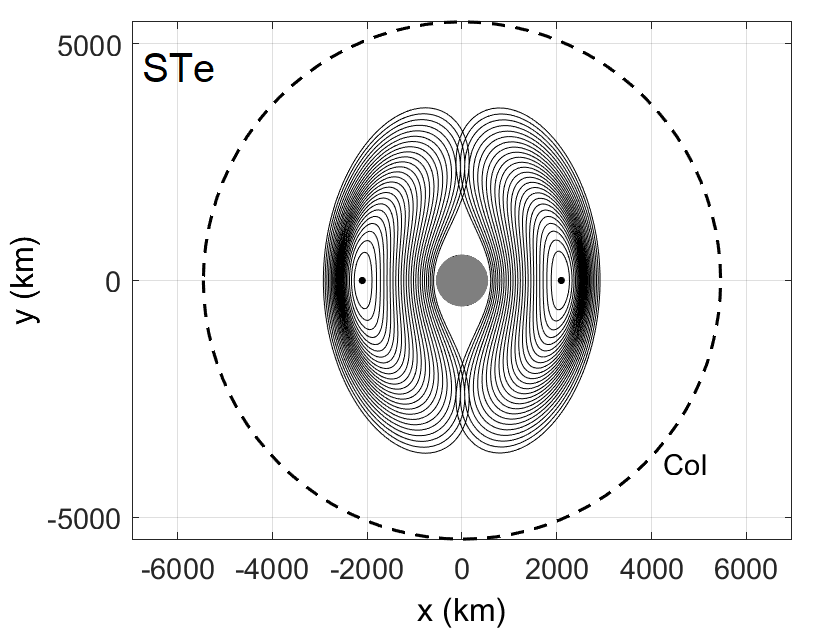} \\
\includegraphics[width=0.49\textwidth]{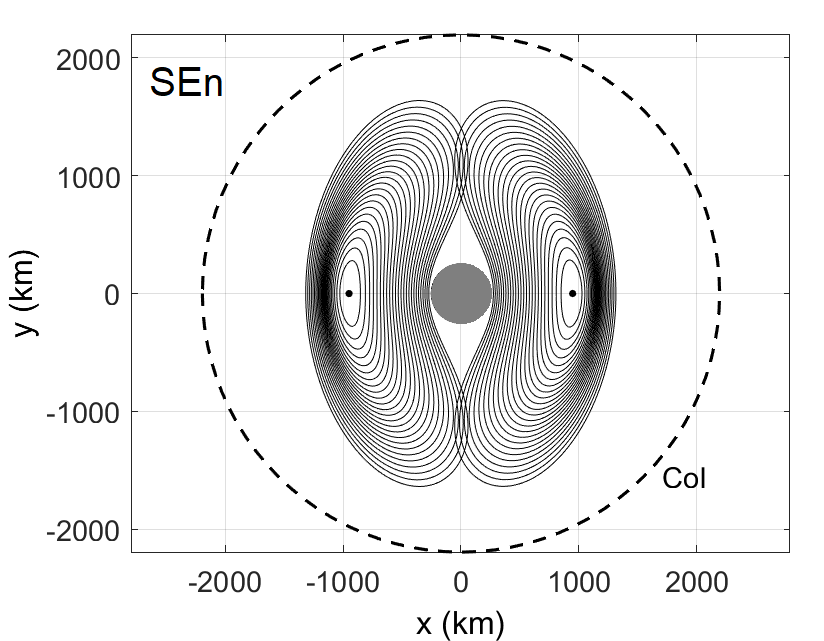} & \includegraphics[width=0.49\textwidth]{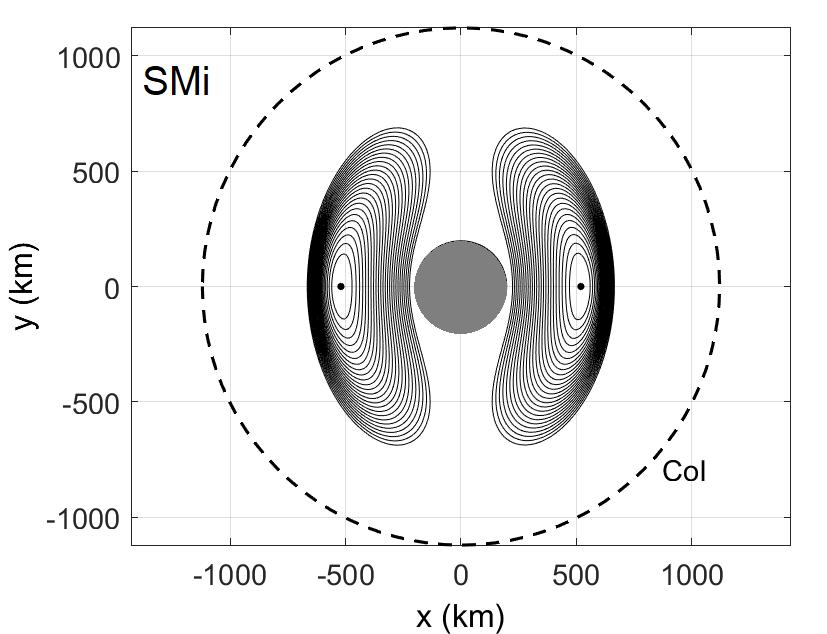} \\
\end{tabular}
\caption{Families of PLOs around L$_1$ and L$_2$ for the four CR3BPs in the respective moon-centered synodic reference frames. CoI is the circle of influence and is defined in Sect.~\ref{sec:inter}.}
\label{fig:lyap}
\end{figure}

Figure~\ref{fig:lyap} illustrates families of planar Lyapunov orbits (PLOs) around L$_1$ and L$_2$ for the four CR3BPs. Each set contains 25 orbits equally spaced in Jacobi constant $C_J$, with the orbits around L$_1$ and L$_2$ having the same $C_J$ value. Individual orbits in the set are identified by an index running from 1 to 25. Table~\ref{tab:families} summarizes the characteristics of the families: minimum and maximum $C_J$ ($C_{J min}$, $C_{J max}$), minimum and maximum period ($P_{min}$, $P_{max}$), minimum and maximum $y$ amplitude ($\Delta y_{min}$, $\Delta y_{max}$), and the orbit index corresponding to the various extrema. 
\begin{table}[h!]
\caption{Main characteristics  of the families of PLOs at L$_1$ and L$_2$ in the four CR3BPs considered: 
minimum and maximum $C_J$ ($C_{J min}$, $C_{J max}$), minimum and maximum period ($P_{min}$, $P_{max}$), minimum and maximum $y$ amplitude ($\Delta y_{min}$, $\Delta y_{max}$) and the orbit index (last row) for which the extrema occur.}
\label{tab:families}
\centering
{\scriptsize \begin{tabular}{llrrrrrr}
\hline
CR3BP & L$_i$		   & $C_{Jmin}$ & $C_{Jmax}$ & $P_{min}$ & $P_{max}$ & $\Delta y_{min}$ & $\Delta y_{max}$    \\ 
     	&	         & -          & -          & (day)     & (day)     & (km)             & (km)  \\ \hline
SMi & L$_1$        & 3.000032   & 3.000068   & 0.457     & 0.548 		 & 288              &	1373	\\     
SMi & L$_2$        & 3.000032   & 3.000068   & 0.458     & 0.549 		 & 281              &	1375	\\ \hline    
SEn & L$_1$        & 3.000036   & 3.000138   & 0.664     & 0.923 	   & 568           & 3273 \\   
SEn & L$_2$        & 3.000036   & 3.000138   & 0.668     & 0.920 	   & 557           & 3275 \\  \hline 
STe & L$_1$        & 3.000114   & 3.000440   & 0.918     & 1.269     & 1240 			   & 7279        \\     
STe & L$_2$        & 3.000114   & 3.000440   & 0.922     & 1.272     & 1190          & 7288    \\  \hline
SDi	& L$_1$        & 3.000097   & 3.000640 	 & 1.324     & 2.038     & 2030          & 13290  \\
SDi	& L$_2$        & 3.000097   & 3.000640   & 1.339 		 & 2.043     & 2080          & 13300 \\ \hline
Orbit index &      & 25         & 1          & 1         & 25        & 1             & 25  \\ \hline
\end{tabular}}
\end{table} 

\begin{figure}[h!]
\centering
\includegraphics[width=0.44\textwidth]{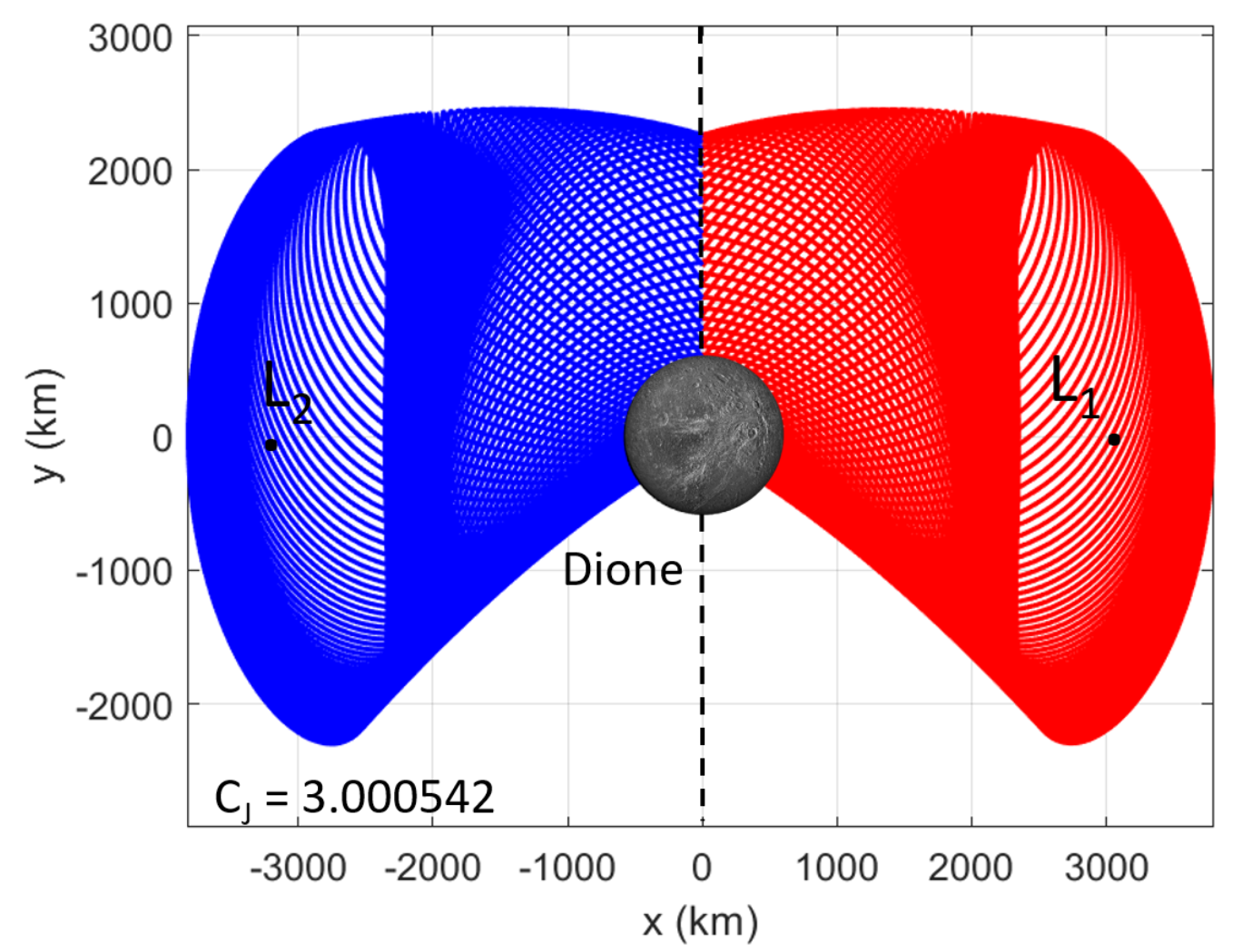} \includegraphics[width=0.43\textwidth]{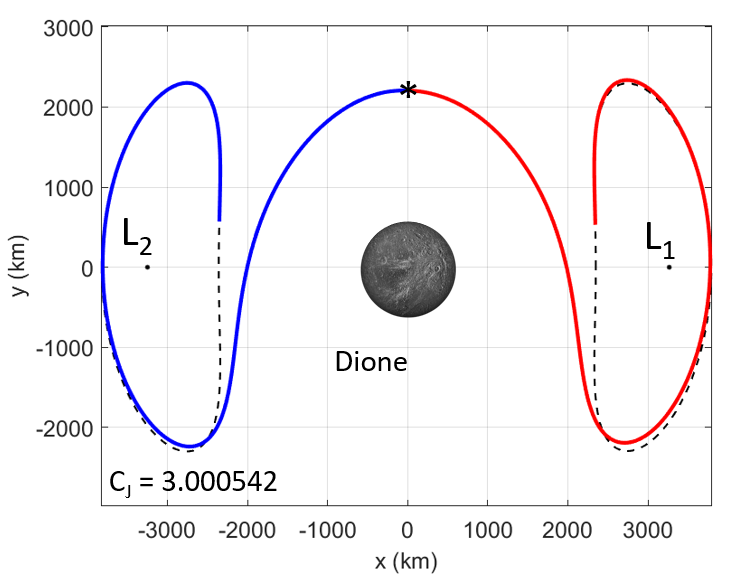} 
\includegraphics[width=0.44\textwidth]{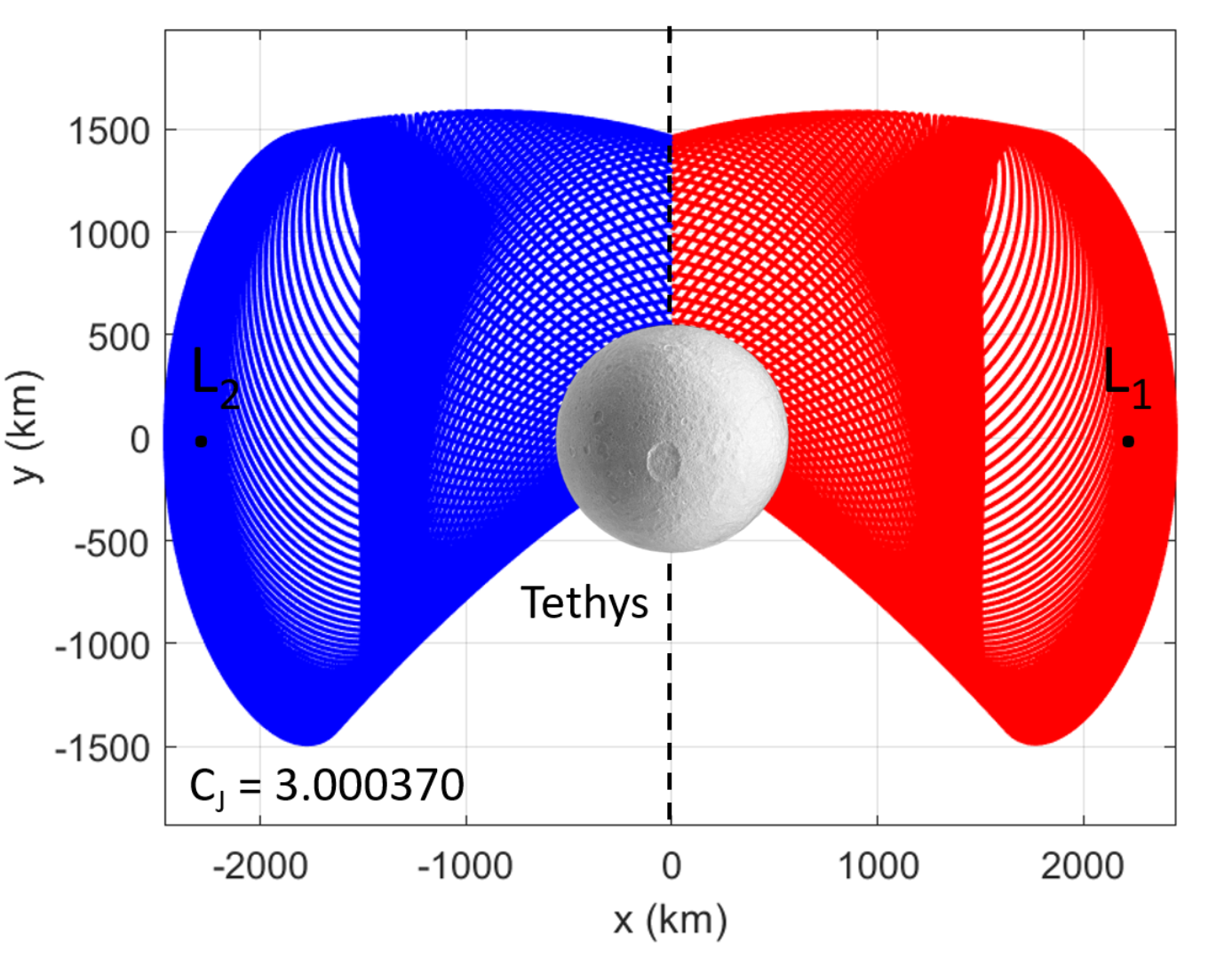} \includegraphics[width=0.43\textwidth]{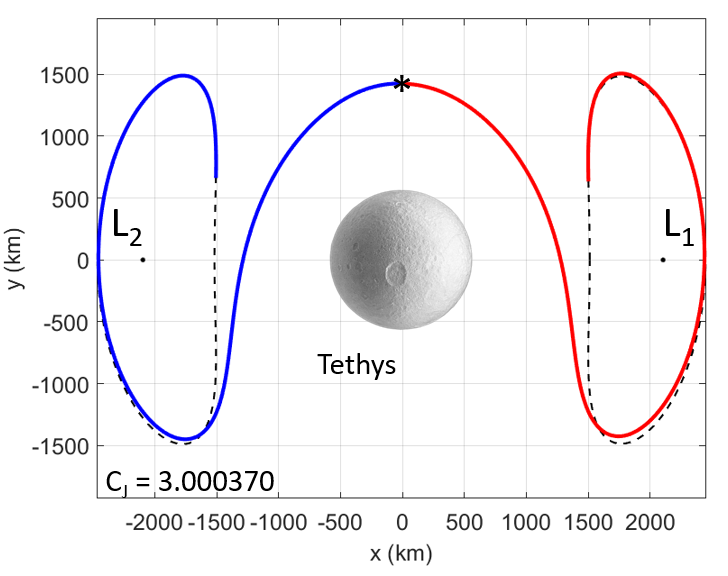} 
\includegraphics[width=0.44\textwidth]{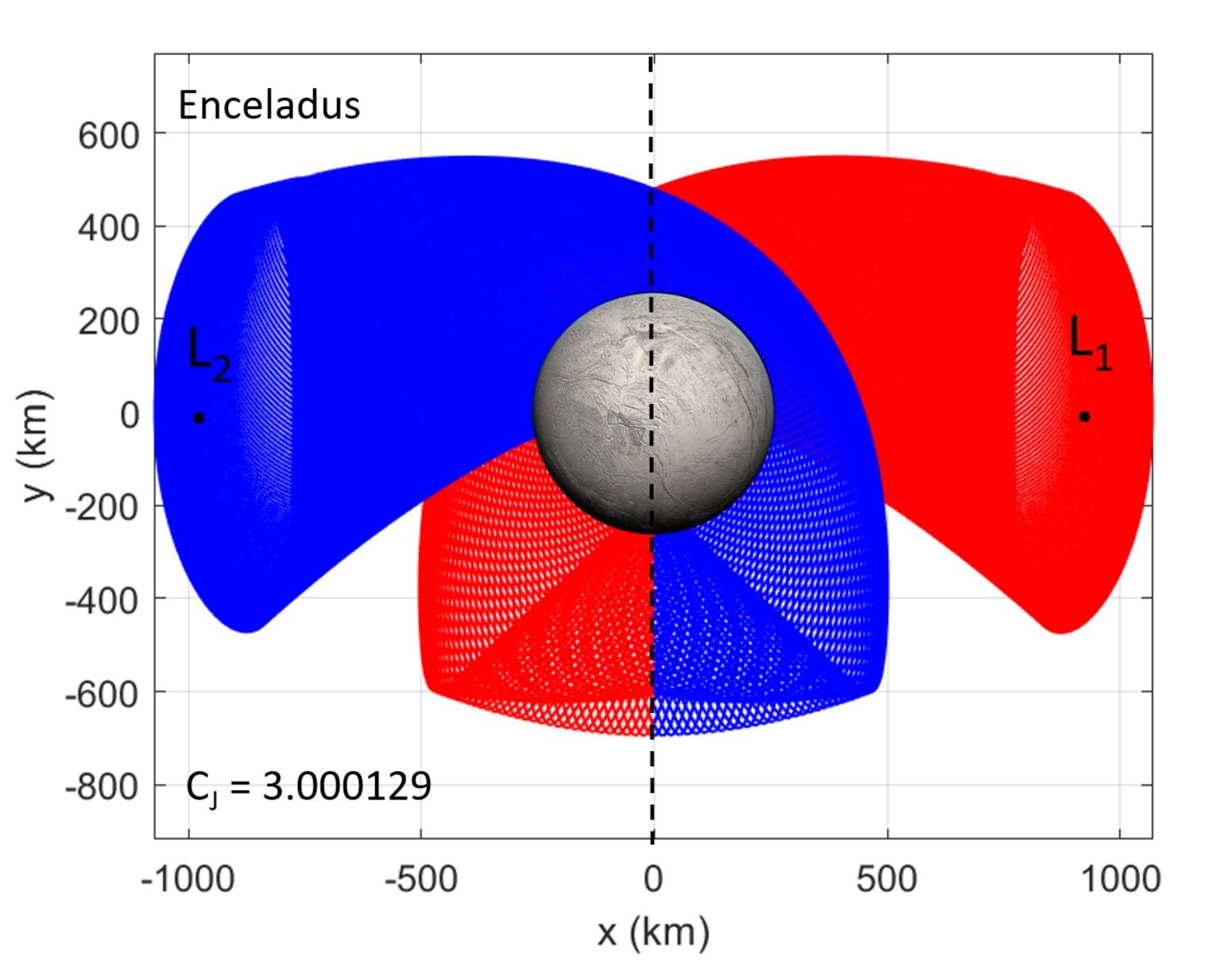} \includegraphics[width=0.44\textwidth]{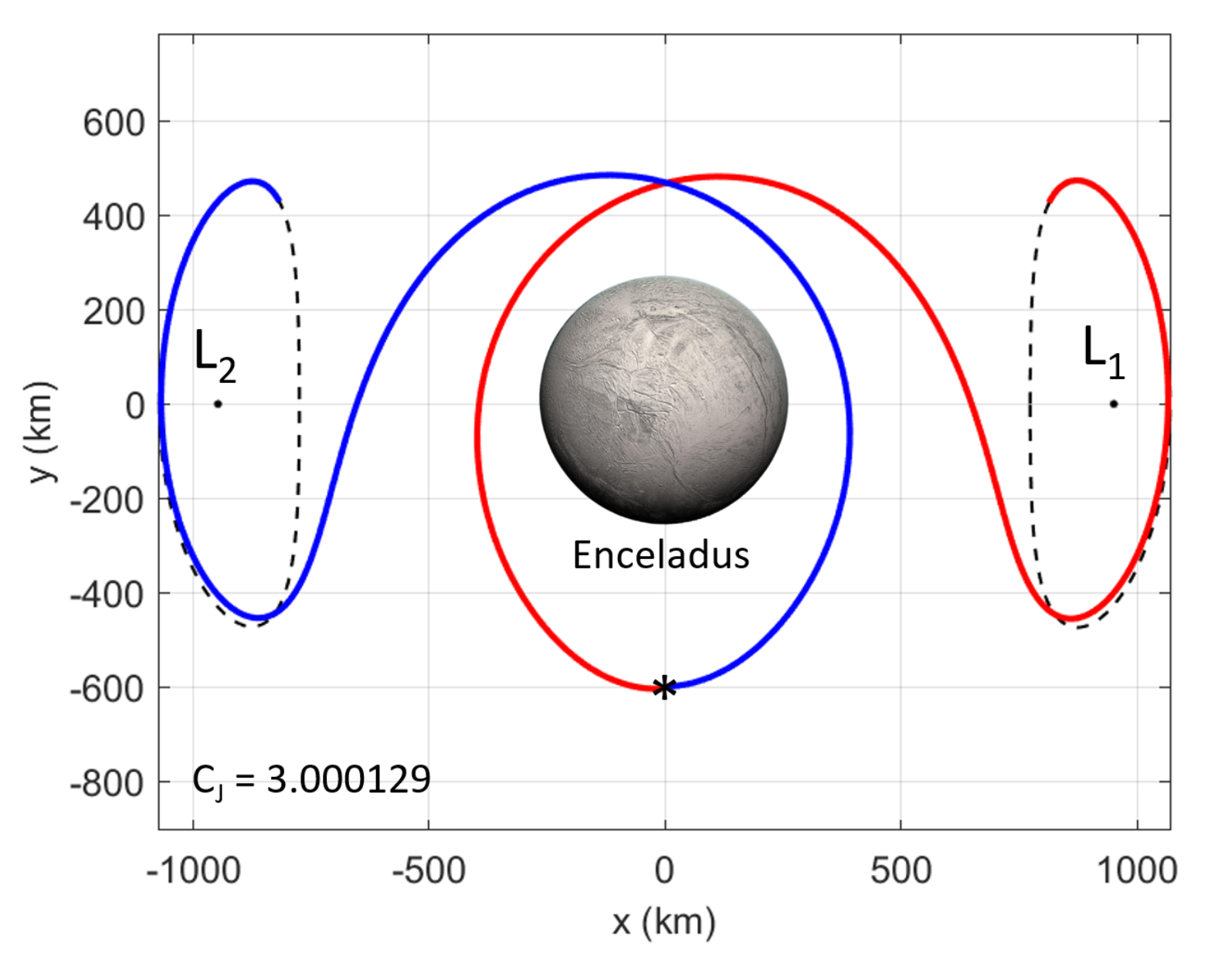} 
\includegraphics[width=0.44\textwidth]{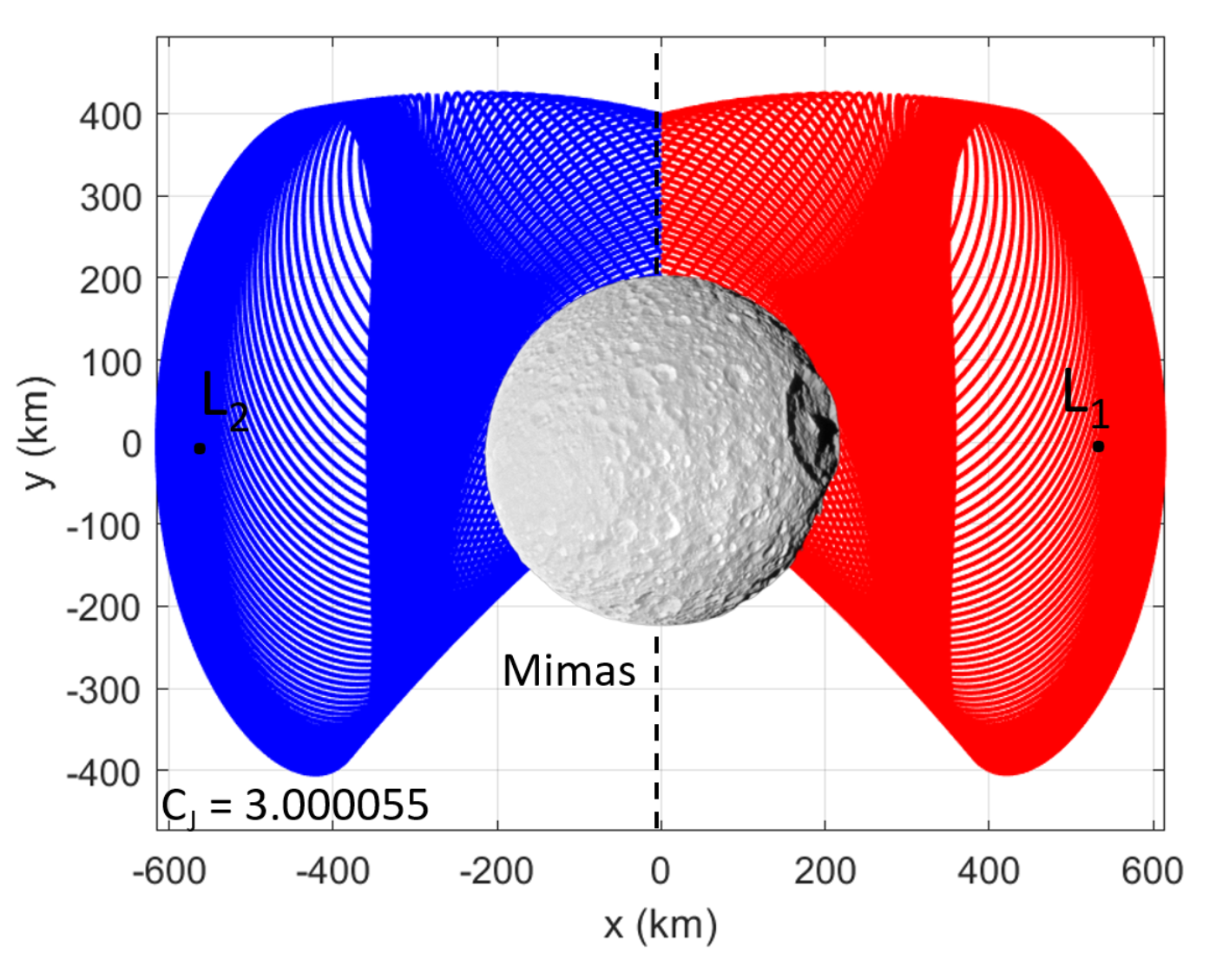} \includegraphics[width=0.44\textwidth]{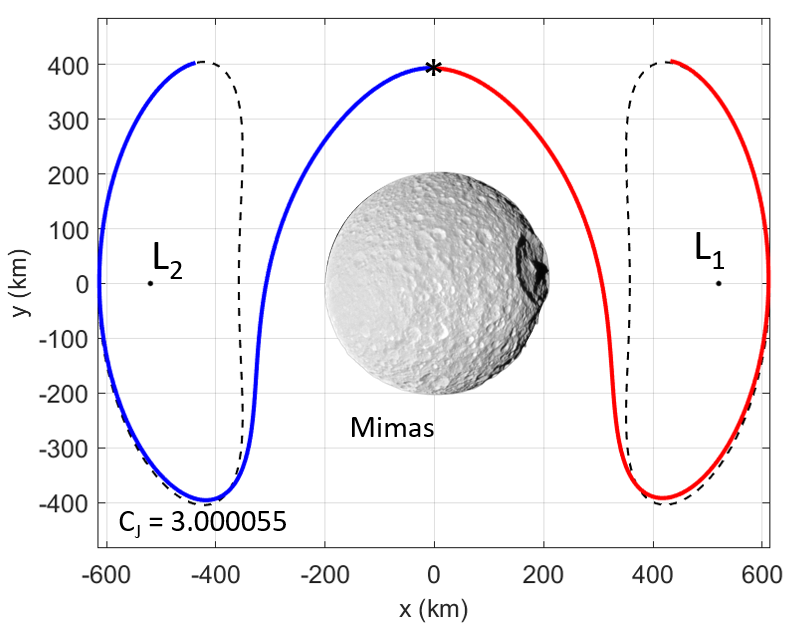} 
\caption{Examples of L$_2$-to-L$_1$ heteroclinic connections in the four CR3BPs: stable and unstable HIMs (left) and connecting arcs (right).}
\label{fig:hetero}
\end{figure}

\begin{figure}[h!]
\centering
\includegraphics[width=0.49\textwidth]{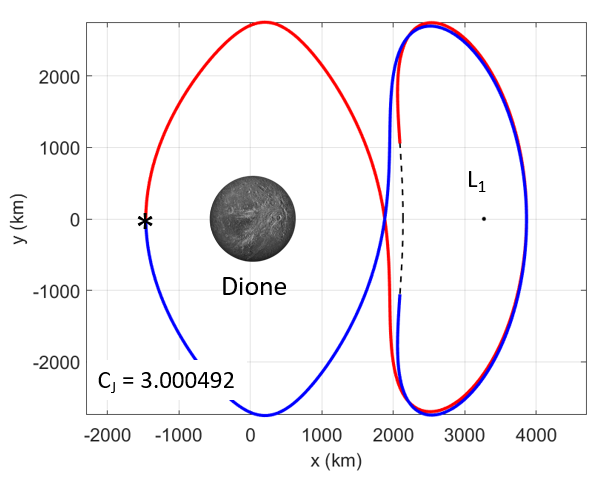} \includegraphics[width=0.49\textwidth]{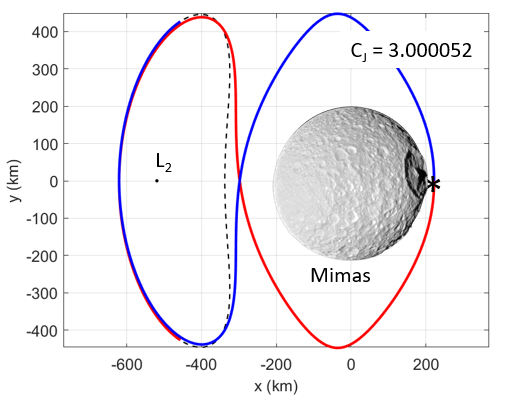} 
\caption{Examples of homoclinic connections in the Saturn-Dione and Saturn-Mimas CR3BPs, respectively around L$_1$ and L$_2$.}
\label{fig:homo}
\end{figure}

The stable and unstable HIMs of the PLOs have been computed and propagated using standard methods, i.e., an initial state is generated by applying a small perturbation in the direction of the stable and unstable eigenvectors of the monodromy matrix of the PLO after appropriate time transformation through the state transition matrix (see e.g., \cite{Parker1989, Zanzottera2012}). Each PLO is discretized with 100 points, each of which corresponds to an invariant manifold trajectory. Then, the branches that develop towards the moon have been used to construct heteroclinic connections between PLOs with the same value of $C_J$ at the two equilibria \cite{Gomez2004,Canalias2006,Barrabes2013,Kumar2021}. The stable and unstable HIMs also serve to construct homoclinic connections, i.e., trajectories that depart and approach the same PLO (see \cite{Canalias2006,Barrabes2009}). By construction, for both types of transfers the cost to leave and approach a PLO is negligible. The stable and unstable segments of the transfer are connected at a suitably defined Poincar\'e section, i.e., the $y$-axis and the $x$-axis of the moon-centered synodic reference frame, respectively, for heteroclinic and homoclinic transfers. For each $C_J$, the transfer is constructed with the arcs for which the magnitude of the velocity difference at the Poincar\'e section is lowest. This minimum impulse is at the level of 1 m/s or less in all cases. Examples of science orbits of the two types around the different moons are depicted in Figs.~\ref{fig:hetero} and \ref{fig:homo}, where the blue and red colors denote stable and unstable HIMs trajectories, respectively. All the computed trajectories satisfy a minimum-altitude constraint of 20 km.
All the heteroclinic transfers are reversible, i.e., symmetric connections exist from L$_2$ to L$_1$. Moreover, the autonomous character of the CR3BP allows to choose the departure time from a PLO arbitrarily. Hence, the same heteroclinic/homoclinic transfer can be used multiple times to explore the same moon. The PLOs can be used as parking orbits between consecutive flights and as gateways to depart the vicinity of a moon and approach the next target in the tour. Additionaly, LT transfers between PLOs of different energy are also possible, as shown for example in \cite{Chupin2017, Zeng2018, Du2022}.

\begin{figure}[h!]
\centering
\includegraphics[width=0.44\textwidth]{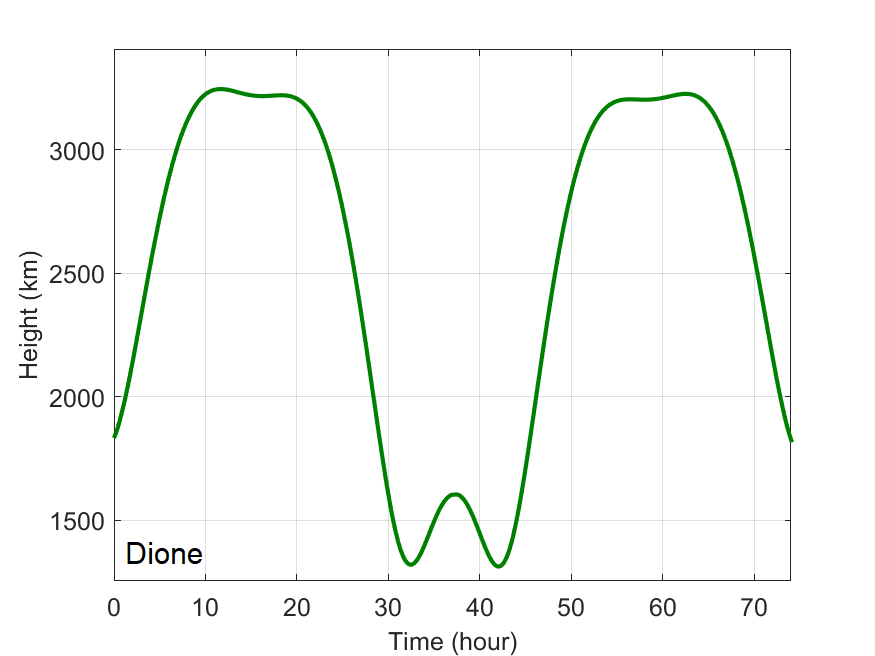} \includegraphics[width=0.44\textwidth]{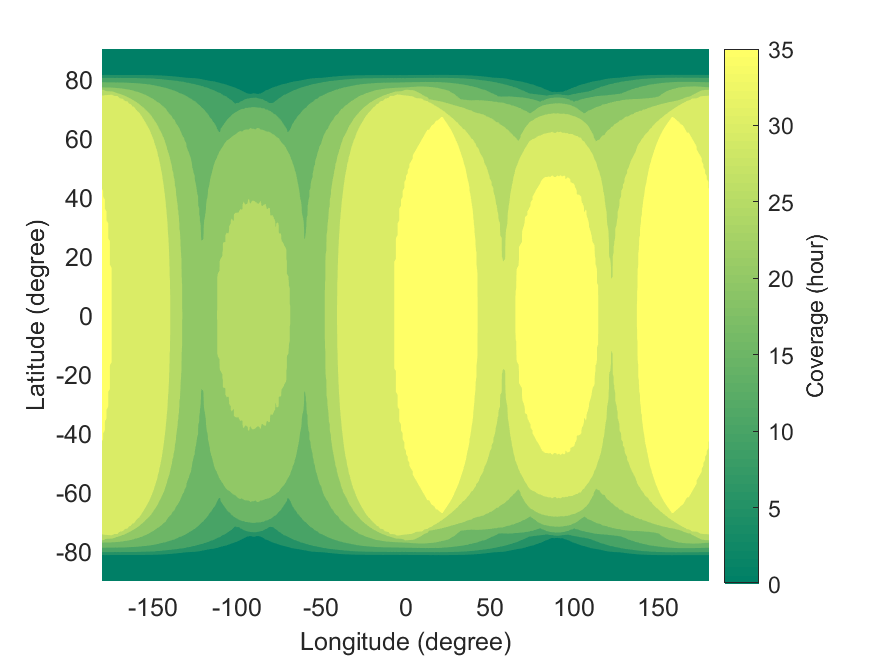} 
\includegraphics[width=0.44\textwidth]{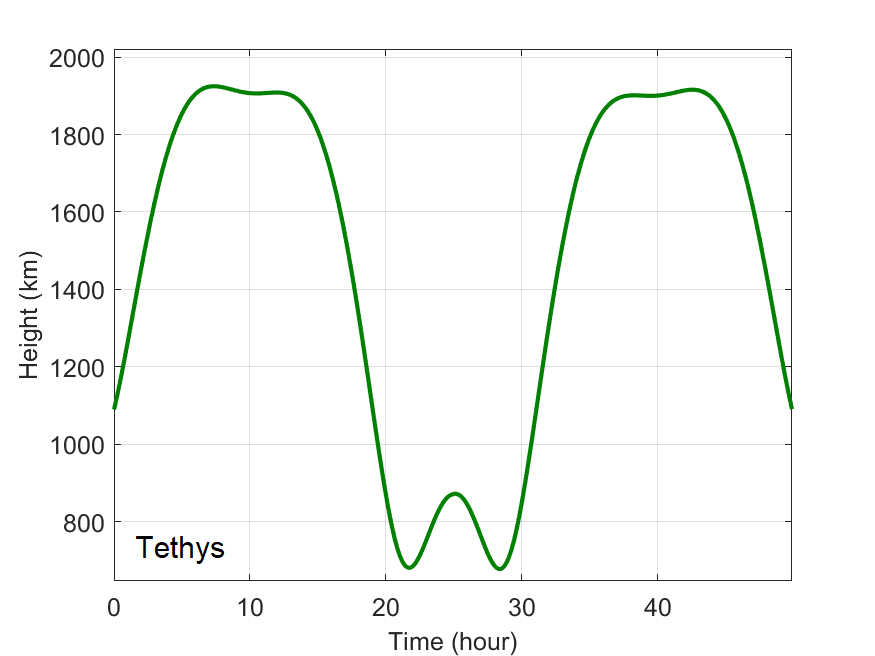} \includegraphics[width=0.44\textwidth]{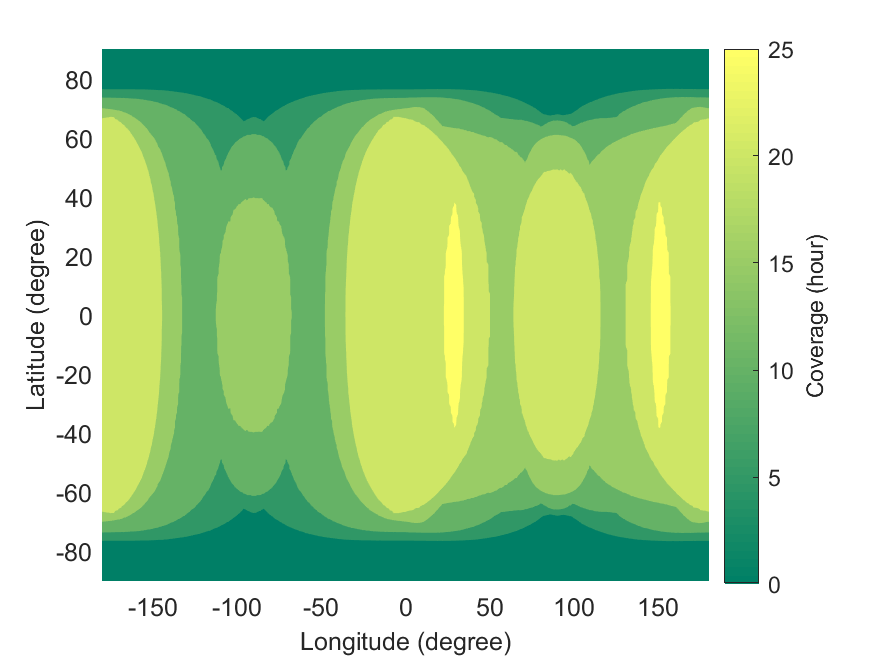} 
\includegraphics[width=0.44\textwidth]{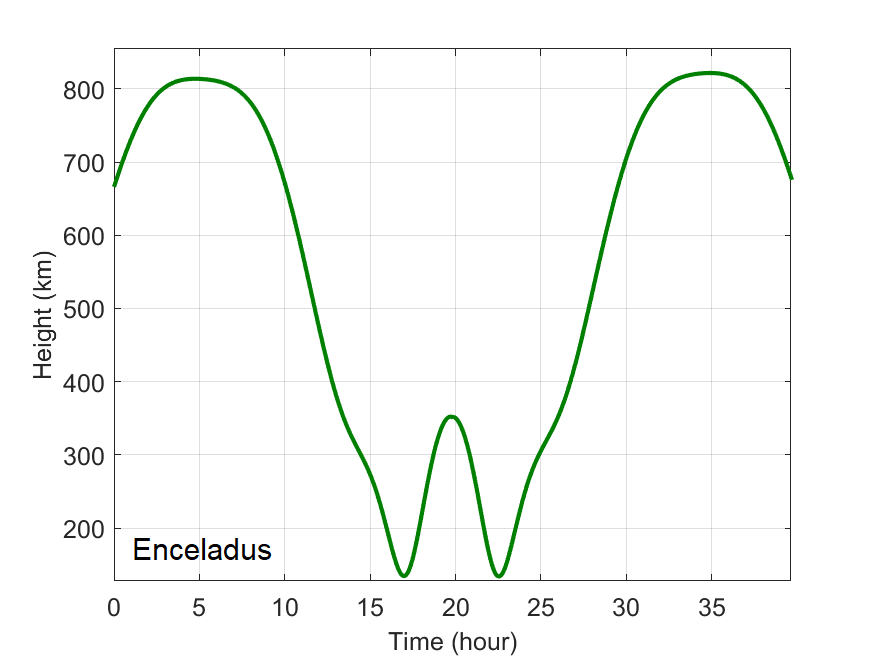} \includegraphics[width=0.44\textwidth]{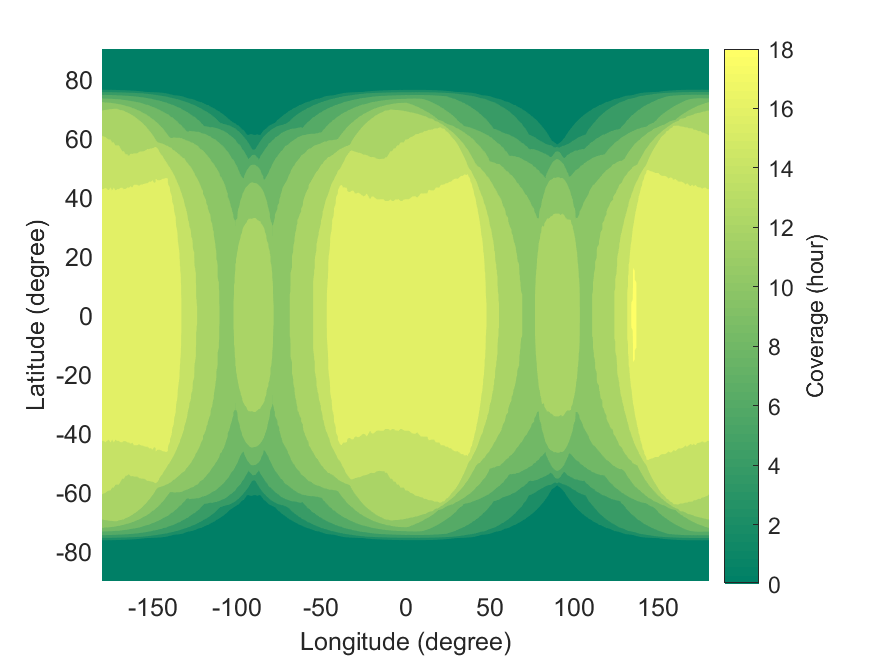} 
\includegraphics[width=0.44\textwidth]{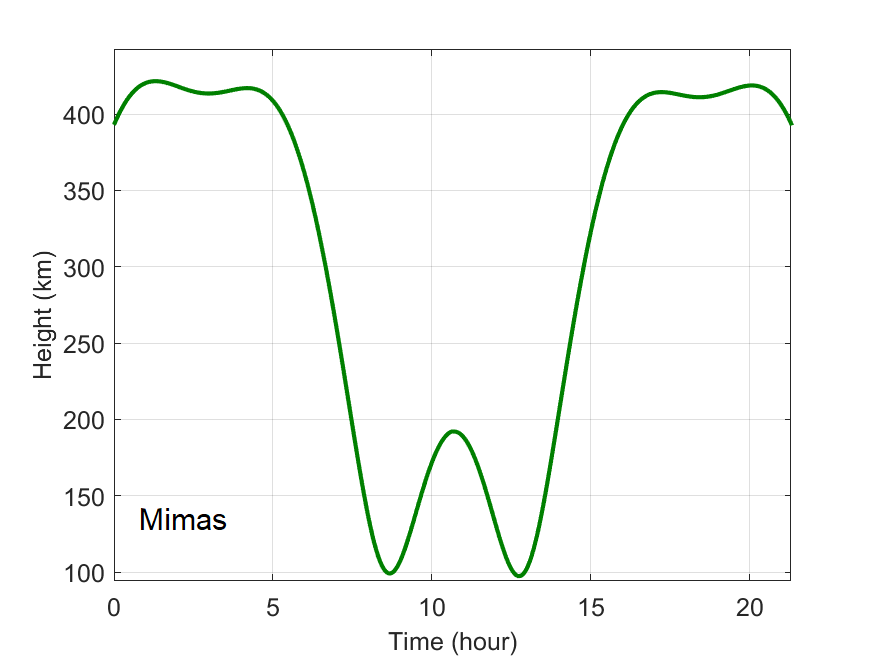} \includegraphics[width=0.44\textwidth]{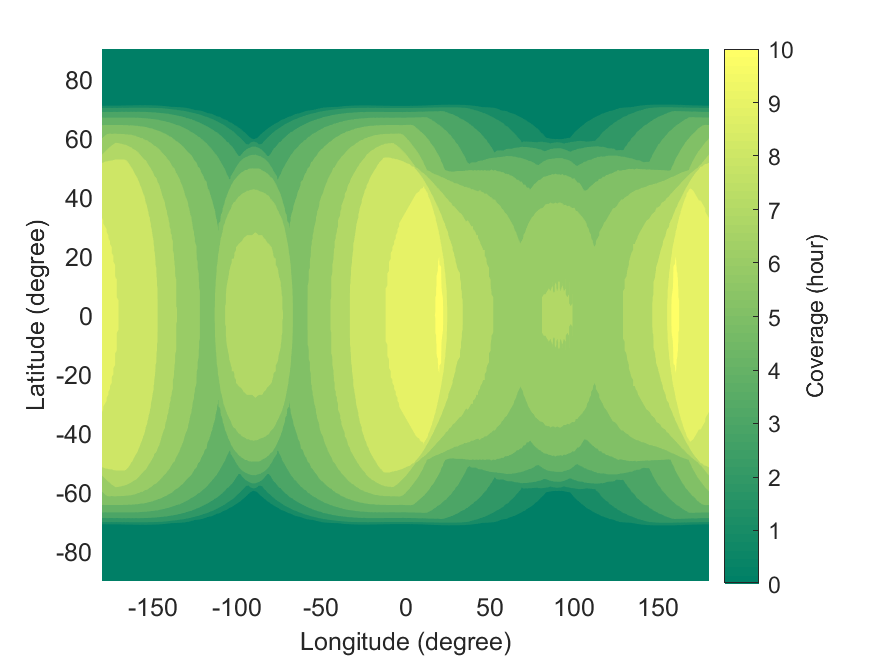} 
\caption{Height above the moon vs. time (left) and surface coverage (right) for the four heteroclinic connections of Fig.~\ref{fig:hetero}.}
\label{fig:hetero_h_c}
\end{figure}

\begin{figure}[h!]
\centering
\includegraphics[width=0.49\textwidth]{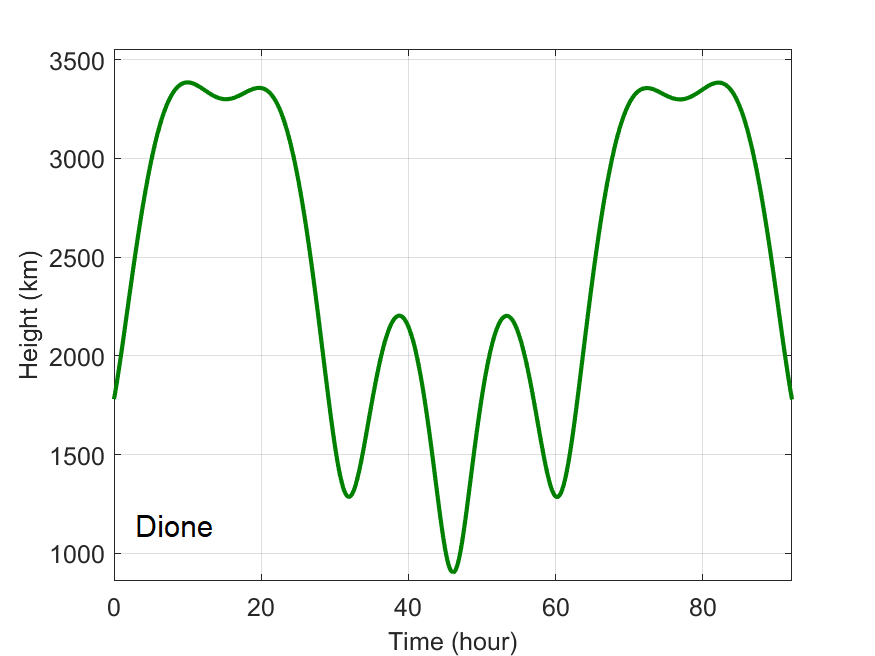} \includegraphics[width=0.49\textwidth]{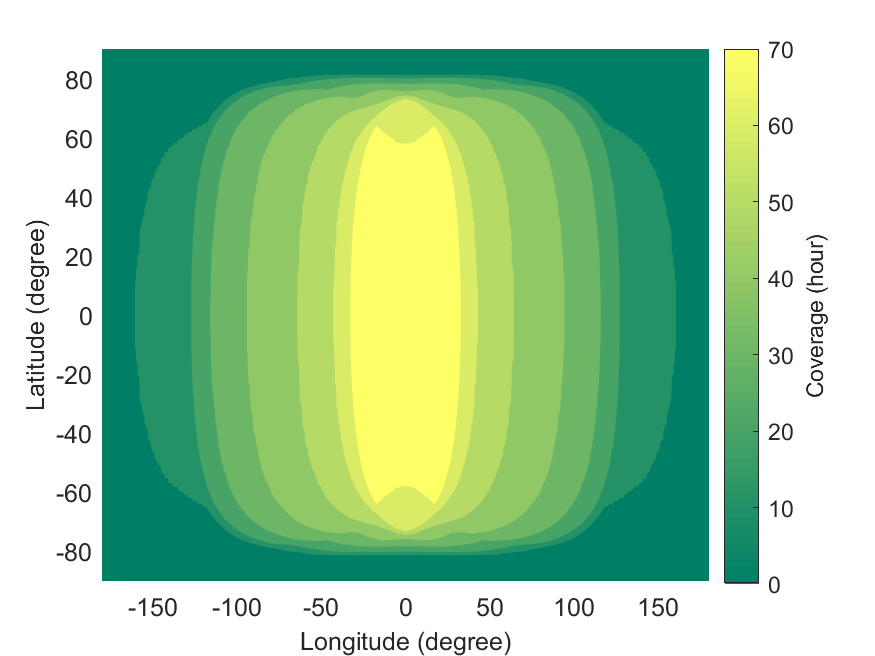} 
\includegraphics[width=0.49\textwidth]{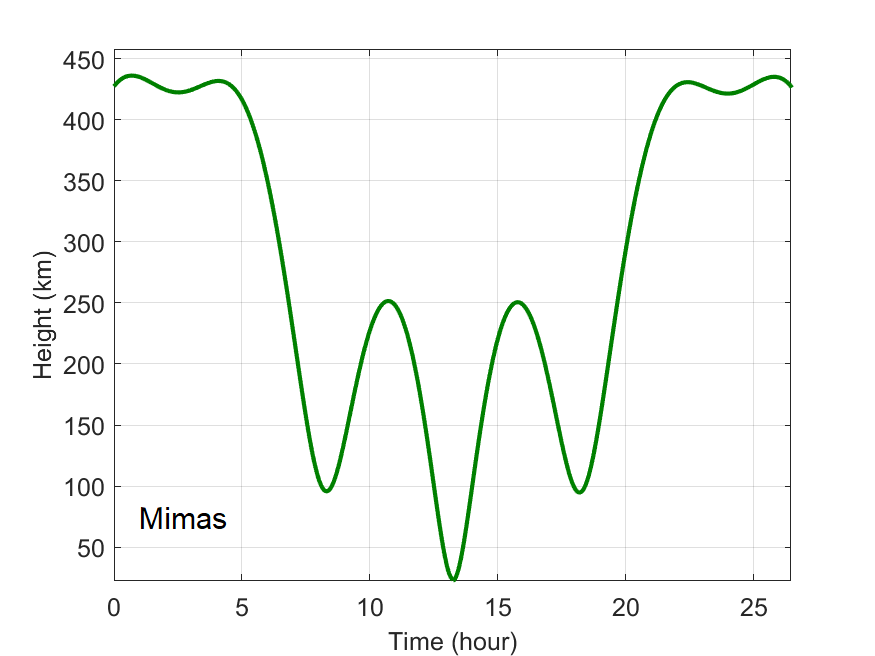} \includegraphics[width=0.49\textwidth]{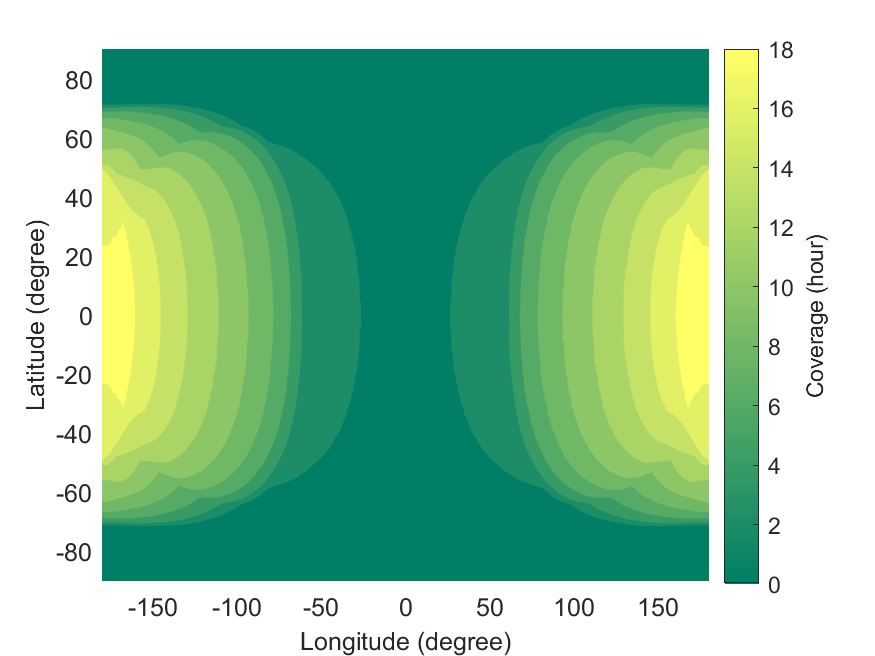} 
\caption{Height above the moon vs. time (left) and surface coverage (right) for the two homoclinic connections of Fig.~\ref{fig:homo}.}
\label{fig:homo_h_c}
\end{figure}

The time history of altitude and the time coverage of the surface for the heteroclinics and homoclinics of Figs.~\ref{fig:hetero} and \ref{fig:homo} are illustrated in Figs.~\ref{fig:hetero_h_c} and \ref{fig:homo_h_c}, respectively. Here, time coverage of a surface element (the discretization employed is $1^{\circ} \times 1^{\circ}$) is the accumulated time during which the S/C is above the local horizon. Note that, due to the condition of tidal lock, the moons do not spin in the respective synodic reference frames. In Figs.~\ref{fig:hetero} and \ref{fig:homo}, the prime meridian is aligned with the positive $x$ axis of the moon-centered synodic reference frame. Depending on the altitude and shape of the trajectory, the time coverage of many areas can reach several tens of hours over individual heteroclinic or homoclinic transfers. In many cases, the maximum visible latitude reaches $80^{\circ}$, meaning that these trajectories allow exploration of a substantial part of the polar regions of the moons.


\subsection{Descent to Dione and inter-moon transfers}
\label{sec:inter}
\begin{figure}[h!]
\centering
\begin{tabular}{lll}
\includegraphics[width=0.27\textwidth]{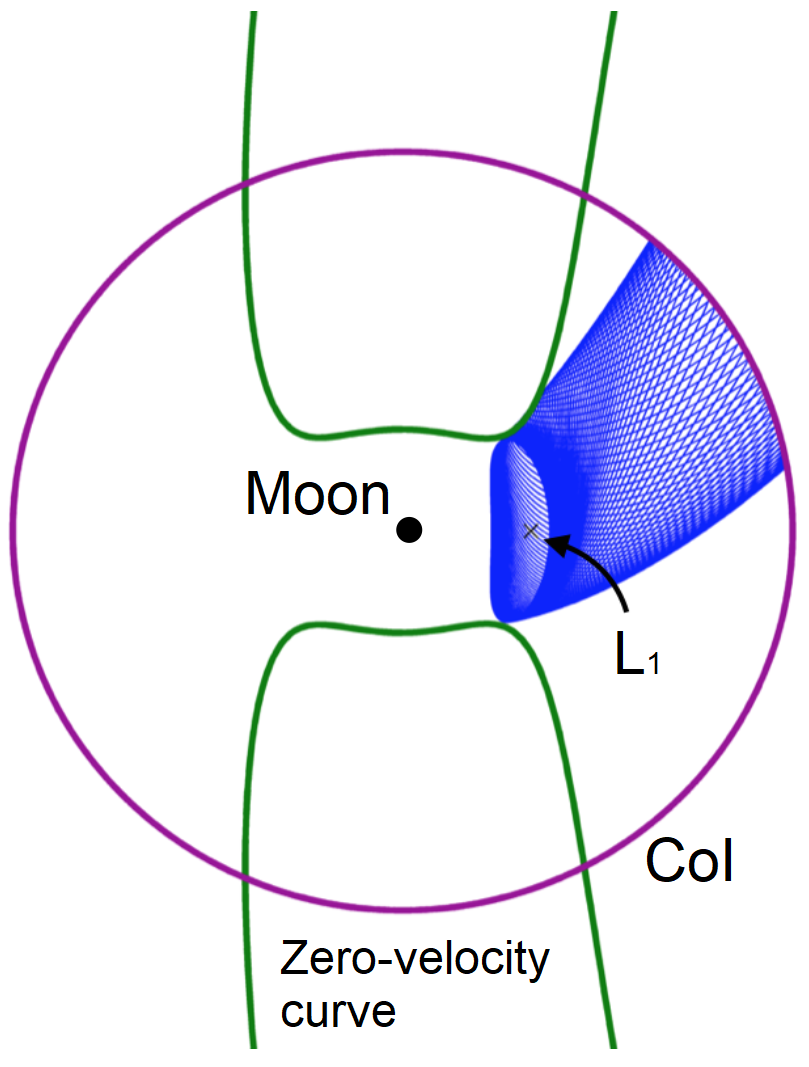} & \hspace{1.0cm} & \includegraphics[width=0.27\textwidth]{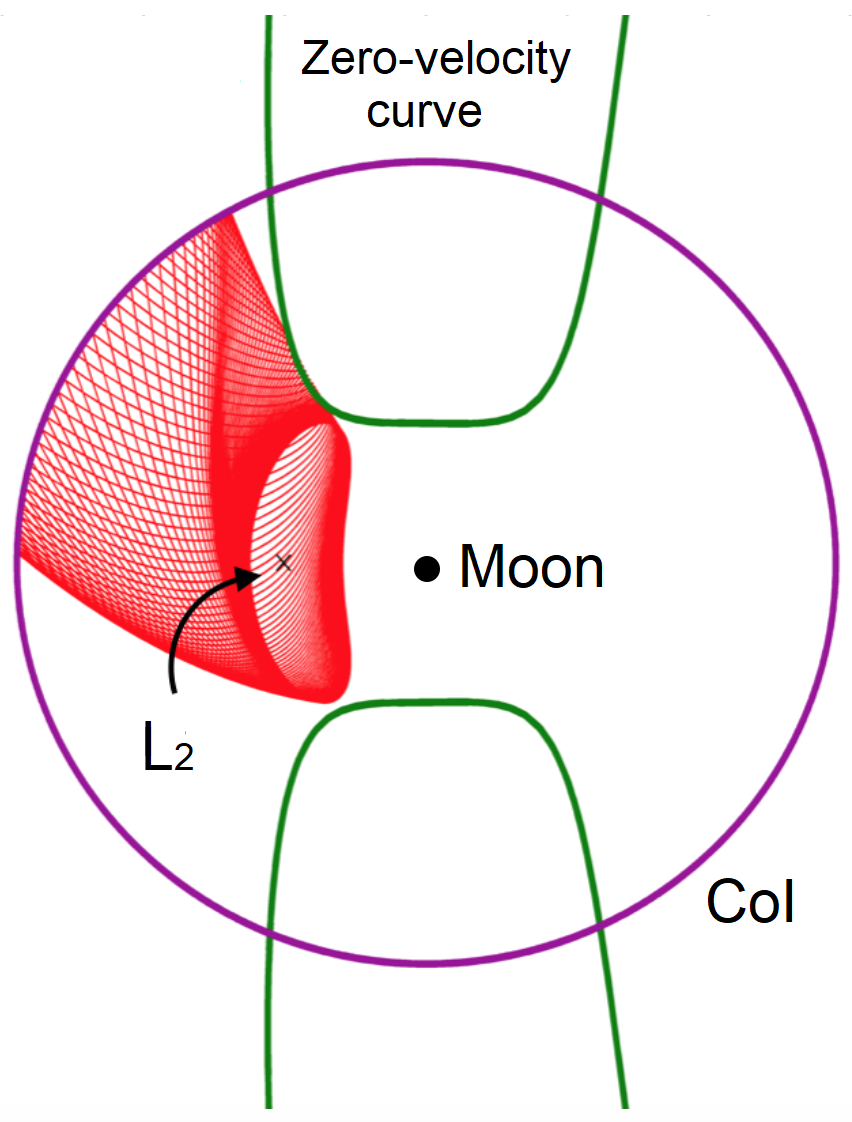} 
\end{tabular}
\caption{Intersections between the outward branches of the stable (left) and unstable (right) HIMs of PLOs around L$_1$ and L$_2$ of a Saturn-moon CR3BP.}
\label{fig:outCoI}
\end{figure}

Following \cite{Fantino2017}, the outward branches of the stable and unstable HIMs (i.e., those that unfold away from the moon, see Fig.~\ref{fig:outCoI}) are propagated until they intersect a moon-centered circle, called circle of influence (CoI), whose radius $r_{CoI}$ equals that of the Laplace sphere of influence \cite{Roy1988} for the given moon scaled by an {\it ad hoc} factor $f \ge 1$:
\begin{equation}
r_{CoI} = f d \left(\frac{m_2}{m_1}\right)^{2/5}.
\end{equation}
$f$ must be sufficiently large to ensure that the CoI encircles all the PLOs of the two families and intersects the outward branches of the HIMs as close to orthogonally as possible (Fig.~\ref{fig:outCoI}).
\begin{figure}[h!]
\centering
\includegraphics[width=0.35\textwidth]{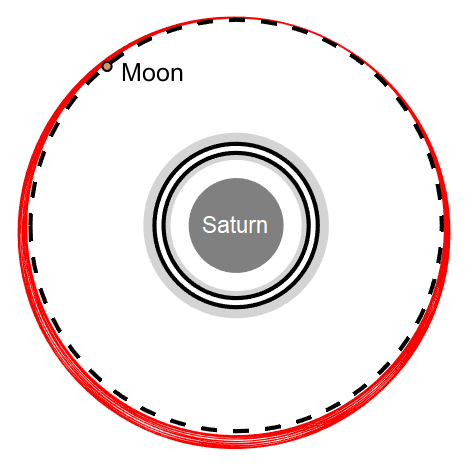}
\caption{Osculating Keplerian Saturn-centered orbits obtained from the states of the unstable HIM trajectories of Fig.~\ref{fig:outCoI} at the CoI of the given moon.}
\label{fig:CoI_HIMs}
\end{figure}
In this study, $f$ has been set equal to 4.5 for all the CR3BPs. Hence, the radii of the four CoIs are 8792, 5457, 2194 and 1121 km for Dione, Tethys, Enceladus and Mimas, respectively (see Fig.~\ref{fig:lyap}). 
The state vectors of the HIMs at the CoI are then represented in the Saturn-centered ICRF and used to compute osculating two-body orbits with focus at Saturn (Fig.~\ref{fig:CoI_HIMs}). Neglecting the gravitational attraction of the moon outside the CoI turns the design of a transfer between moons in adjacent orbits into the search for intersections between coplanar, confocal ellipses, as shown in \cite{Fantino2017} for the case of the Galilean moons. The velocity difference between the connecting arcs at their intersection point(s) can be compensated for by an impulsive maneuver\footnote{\cite{Fantino2017} showed that in the case of the Galilean moons, disregarding the gravitational attraction of the moon outside the CoI introduces an error of only 0.6\% in the magnitude of the impulse at the point of intersection.}.

\begin{figure}[h!]
\centering
\includegraphics[width=0.65\textwidth]{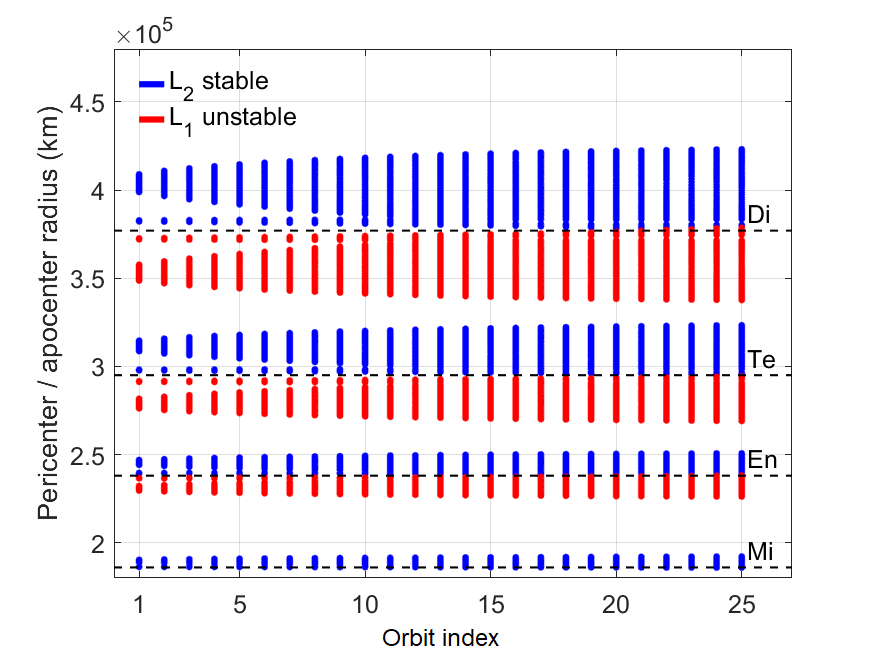} 
\caption{Pericenter and apocenter radii of the osculating Keplerian orbits corresponding to stable and unstable HIM trajectories emanating from PLOs around L$_1$/L$_2$ of the four CR3BPs.}
\label{fig:peri_apo}
\end{figure}

\begin{figure}[h!]
\centering
\includegraphics[width=0.65\textwidth]{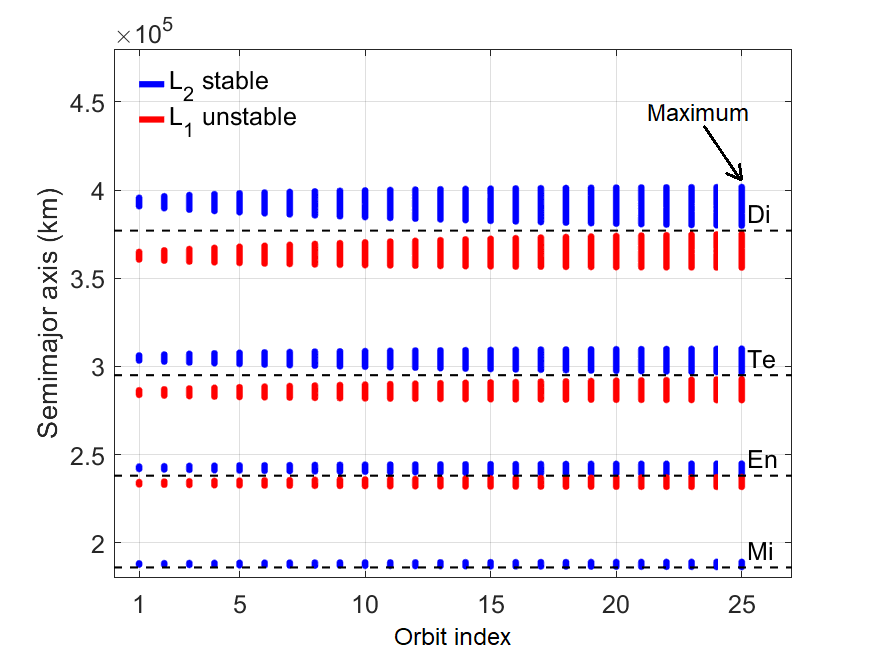} 
\caption{Semimajor axes of the osculating Keplerian orbits corresponding to stable and unstable HIM trajectories emanating from PLOs around L$_1$/L$_2$ of the four CR3BPs. The orbit approaching Dione through L$_2$ with the largest semimajor axis and indicated with an arrow is the target of the descent from Titan to Dione.}
\label{fig:semimajor}
\end{figure}

\begin{figure}[h!]
\centering
\includegraphics[width=0.495\textwidth]{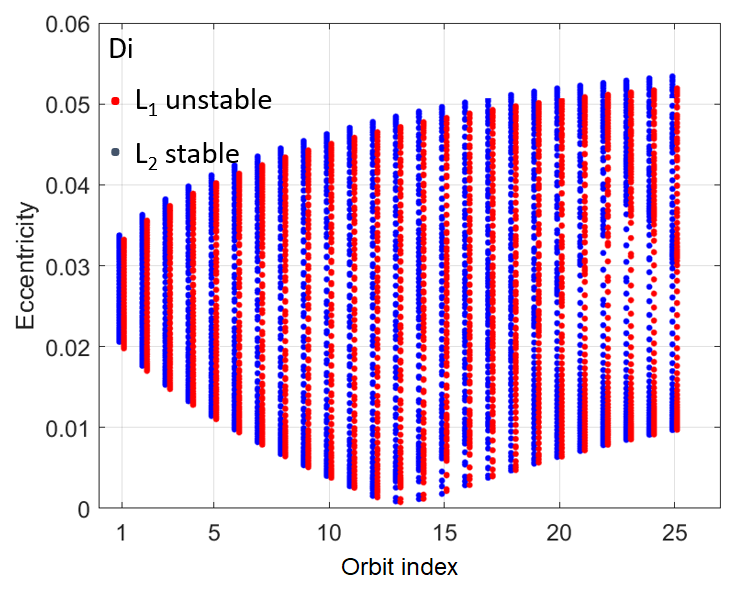} \includegraphics[width=0.495\textwidth]{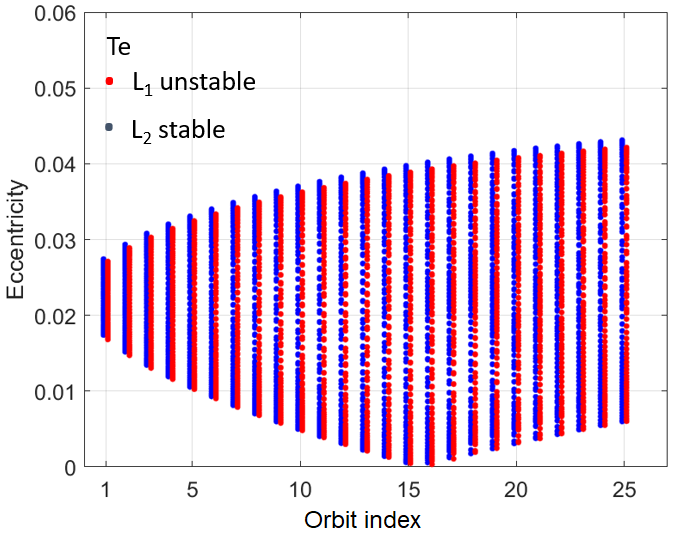} \\
\includegraphics[width=0.495\textwidth]{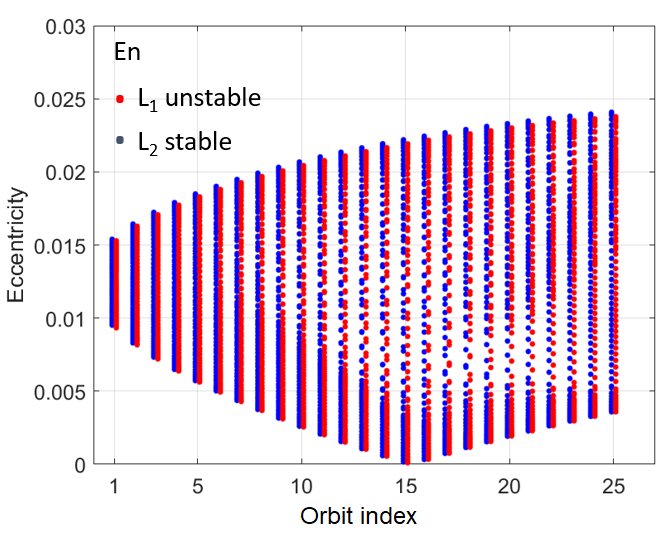} \includegraphics[width=0.495\textwidth]{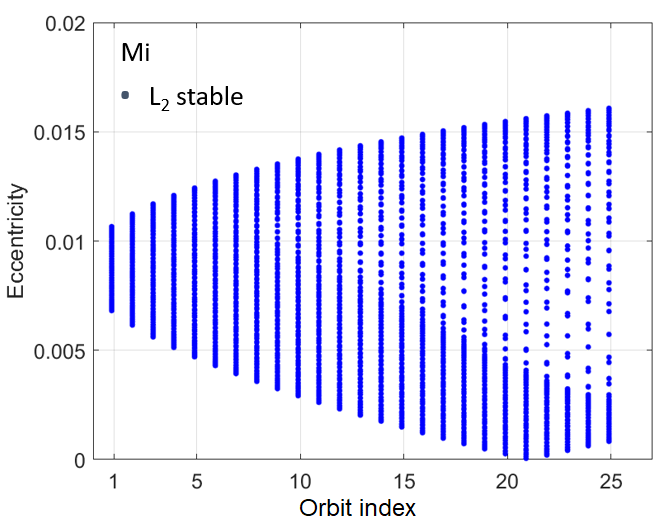} 
\caption{Eccentricities of the osculating Keplerian orbits corresponding to stable and unstable HIM trajectories emanating from PLOs around L$_1$/L$_2$ of the four CR3BPs.}
\label{fig:eccen}
\end{figure}

In the planar approximation, the orbital elements of the osculating ellipses are the semimajor axis $a$, the eccentricity $e$ and the longitude of the pericenter $\omega$. 
In the case of the ILMs, the osculating ellipses emanating from adjacent moons do not intersect. Their eccentricities are very small, while the inter-moon distances are large. This can be seen in Fig.~\ref{fig:peri_apo} which depicts the pericenter and apocenter radii of all the ellipses generated from the four CR3BPs as functions of the orbit index, with different colors for stable and unstable trajectories and  black dashed lines representing the orbital radii of the four moons. In the case of Mimas, only solutions corresponding to L$_2$ are shown because this is the innermost moon in the tour and no transfers departing the vicinity of its L$_1$ point are considered. For the other three moons in the itinerary, i.e., Dione, Tethys and Enceladus, separate sets of values appear for orbits emanating from L$_1$ and L$_2$. The lack of overlap between ranges associated with distinct moons proves that direct impulsive transfers between ILMs are not possible. Figures~\ref{fig:semimajor} and \ref{fig:eccen} illustrate the orbital semimajor axes and eccentricities of all the osculating ellipses. 
\begin{figure}[h!]
\centering
\includegraphics[width=0.50\textwidth]{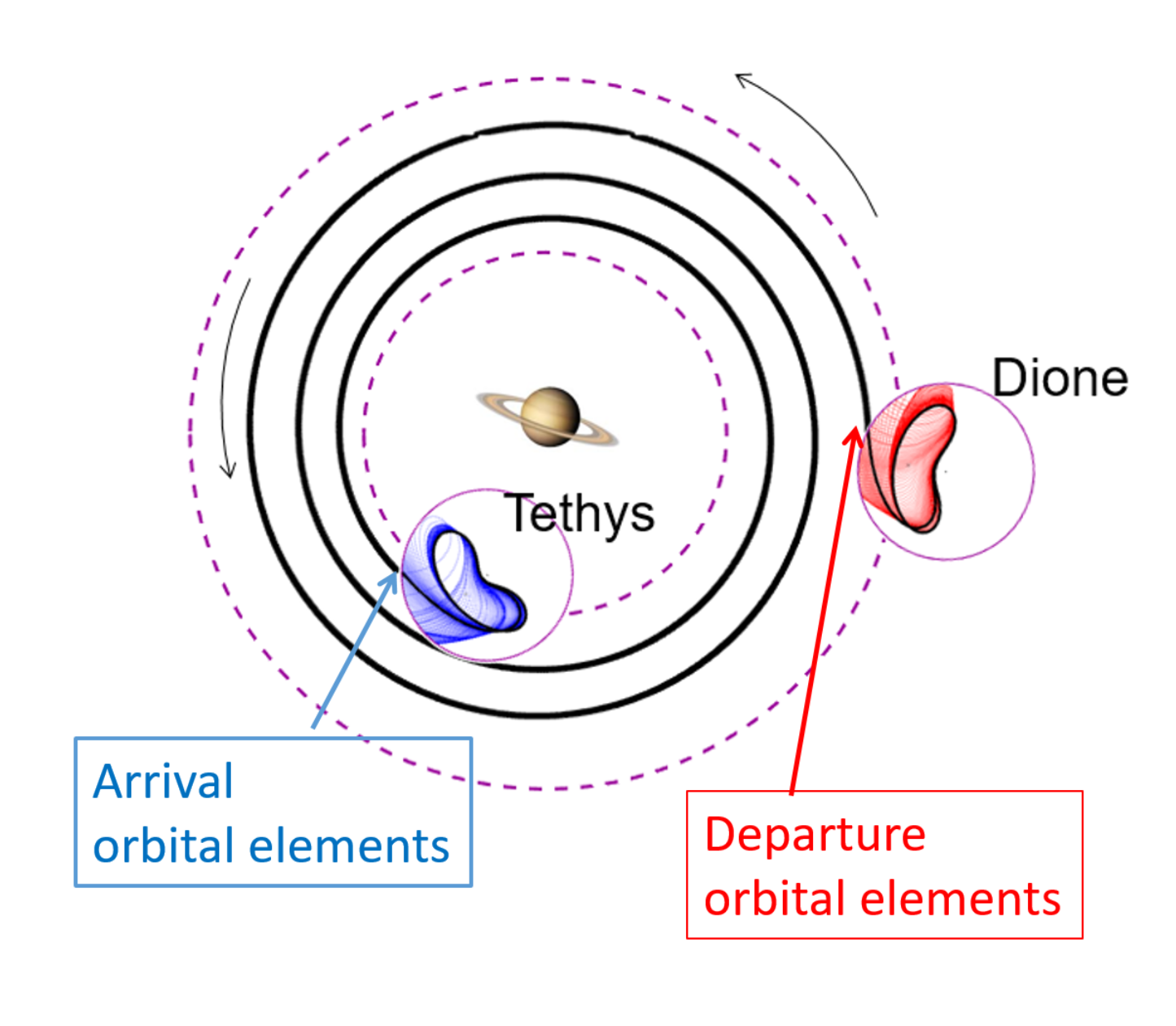} 
\caption{The inter-moon transfer (here from Dione to Tethys) is an LT trajectory between osculating orbits originating from the PLOs of the respective CR3BPs.}
\label{fig:Di2Te}
\end{figure}

The approach adopted in this work to design inter-moon connections is based on LT transfers. The concept is sketched in Fig.~\ref{fig:Di2Te} for a trajectory from a PLO around the L$_1$ point of Dione to a PLO around the L$_2$ point of Tethys. 
Moon-to-moon transfers in the tour always originate in the vicinity of the L$_1$ point of the outer moon and approach the L$_2$ point of the inner one. The direction of the transfer determines the stability character of the HIMs at each end, i.e., unstable at the outer moon, stable at the inner moon. The propelled arc starts and ends at the CoIs of the departure and arrival moons. 
In the case of the descent from Titan to Dione, the LT transfer aims to connect the last post-GA orbit (see Table~\ref{tab:flybys}) and the osculating ellipses corresponding to the stable manifolds of PLOs around L$_2$ in the Saturn-Dione CR3BP. 
In all cases, uninterrupted thrust at 36 mN (specific impulse of 1600 s) is applied in a direction determined locally by a guidance law that maximizes the instantaneous rate of reduction of an error function 
$\Im$ expressed in terms of the osculating orbital elements.  
The strategy aims at modifying the departure elements ($a$, $e$) and 
making them coincide with those ($\bar{a}$, $\bar{e}$) of the arrival orbit. The form chosen for the error function $\Im$ is
\begin{equation}
\Im = (a-\bar{a})^2 + \left(ae-\bar{ae}\right)^2, 
\end{equation}
in which the eccentricity is multiplied be the semimajor axis to keep the expression dimensionally homogeneous and avoid large differences in the order of magnitude of the two components of the error when natural units are not used.
The typical time scale of this LT transfer is at least one order of magnitude longer the than orbital periods of the moons. Therefore, the phasing requirements can be ignored, since the departure time can be
adjusted suitably without a significant effect on the transfer duration. In fact, it is even possible to use an arbitrary departure time and correct the phase a posteriori, with a slight reduction of the thrust magnitude to introduce the appropriate delay.
With this feature in mind, the error function is assumed to depend only
on the semimajor axis and eccentricity, i.e., $\Im(a,e)$. This is not a requirement of the method, but it helps in keeping the expressions simple.
What follows is a direct application of the classical gradient descent optimization technique \cite{Cauchy1847}. 
In the context of modern guidance algorithms, it can be considered a particular case of the general Proximity Quotient guidance law (Q-Law) \cite{Petropoulos2004,SeungwonB2005,Falck2014}. 
Details of the algorithm can be found in \cite{Fantino2020}. Note that since the thrust is constant, the time-optimal transfers are also the most efficient in terms of propellant consumption.

\begin{figure}[h!]
\centering
\includegraphics[width=0.48\textwidth]{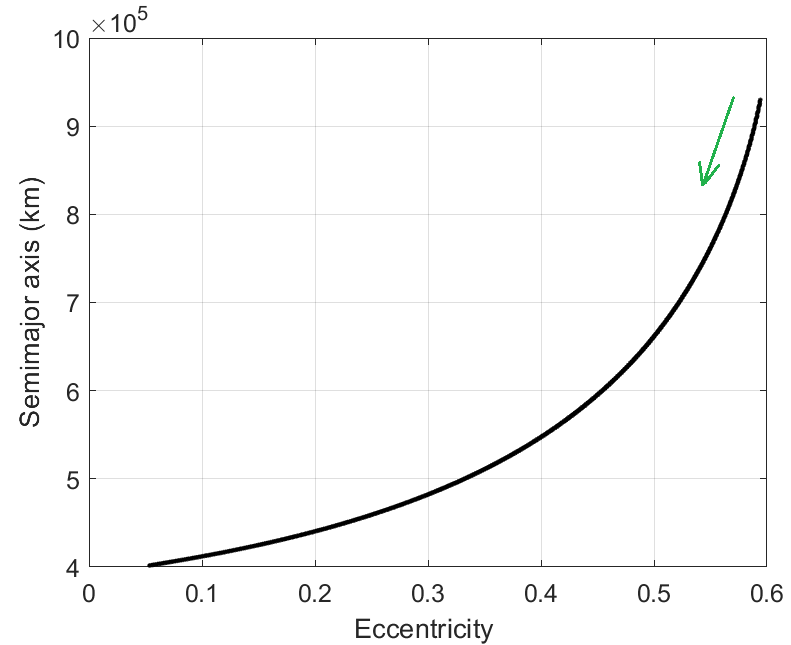}
\includegraphics[width=0.48\textwidth]{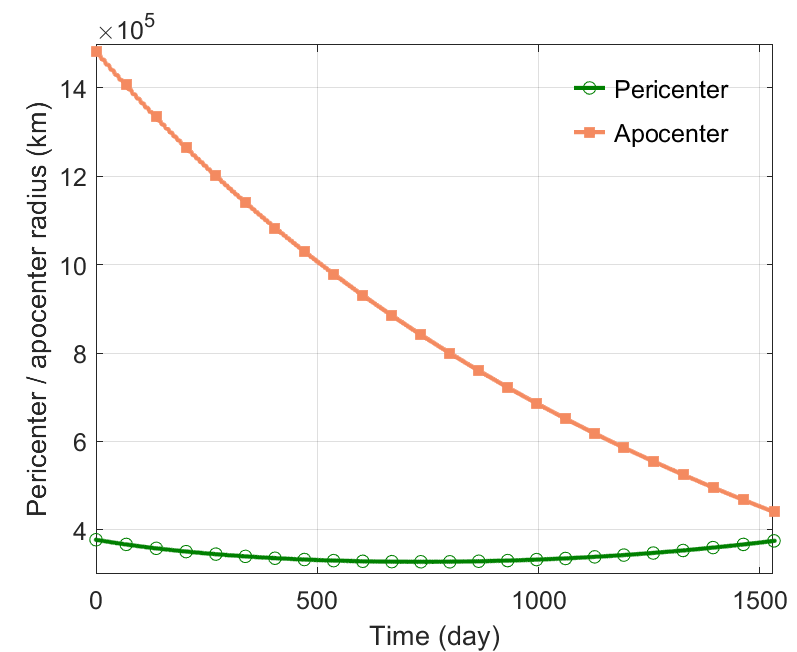} 
\caption{Variation of semimajor axis and eccentricity (left), pericenter and apocenter radius (right) over the optimal LT transfer from Titan to Dione.}
\label{fig:a_e}
\end{figure}

\begin{figure}[h!]
\centering
\includegraphics[width=0.50\textwidth]{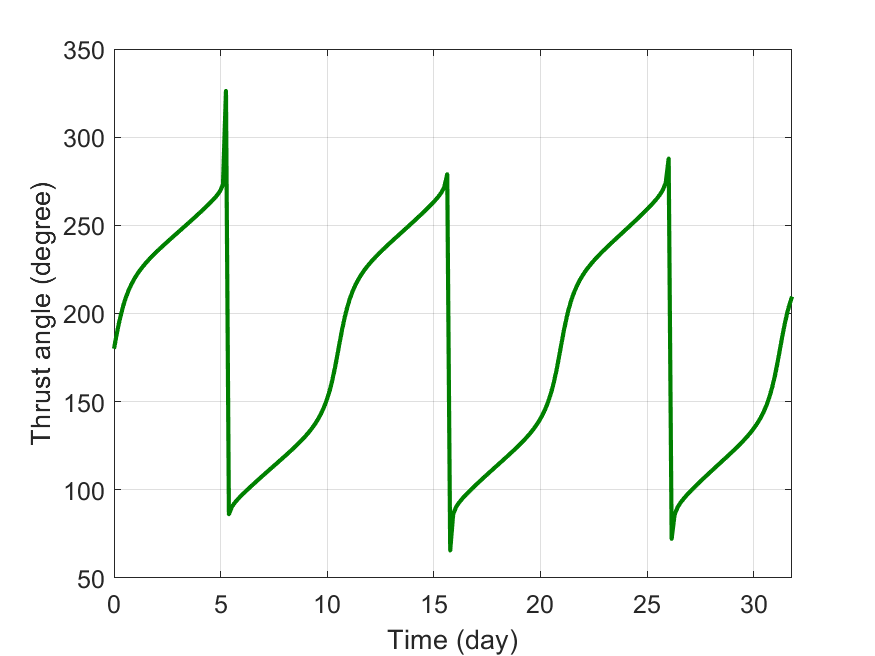}
\caption{Thrust angle history over the first three revolutions around Saturn during the LT transfer from Titan to Dione.}
\label{fig:th_ang}
\end{figure}

For the descent from Dione to Titan, there is one departure orbit and $100 \times 25$ possible sets of arrival orbital elements. The optimal (i.e., minimum-$\Delta V$) trajectory connects to the PLO with index 25 of Saturn-Dione CR3BP. The arrival osculating ellipse has the largest semimajor axis in the set (see Fig.~\ref{fig:semimajor}) and, consequently, the smallest energy difference with respect to the last post-GA orbit. This highlights the importance of the energy difference in the cost of the transfer. However, energy is not the only factor to considere, eccentricity also plays a role. This is demonstrated by the inter-moon transfers (see next paragraph) where the most efficient solution is not always between orbits with the minimum energy gap. For the Dione to Titan transfer, the S/C performs 290 revolutions around Saturn. The time of flight is 1574 days, the propellant consumption is 312 kg and the velocity variation is 5770 m/s. The mass of the S/C upon arrival at Dione is 702 kg. As indicated at the beginning of Sect.~\ref{sec:tour}, the transfer time could be improved performing GAs with Rhea during the descent.
Figure~\ref{fig:a_e} shows the variation of semimajor axis (from $9.3 \cdot 10^5$ km to $4.0 \cdot 10^5$ km),  eccentricity (from 0.5943 to 0.0535), pericenter and apocenter radii over the transfer. Figure~\ref{fig:th_ang} illustrates the thrust angle (measured clockwise from the circumferential direction) over the first three revolutions. 

\begin{figure}[h!]
\centering
\includegraphics[width=0.62\textwidth]{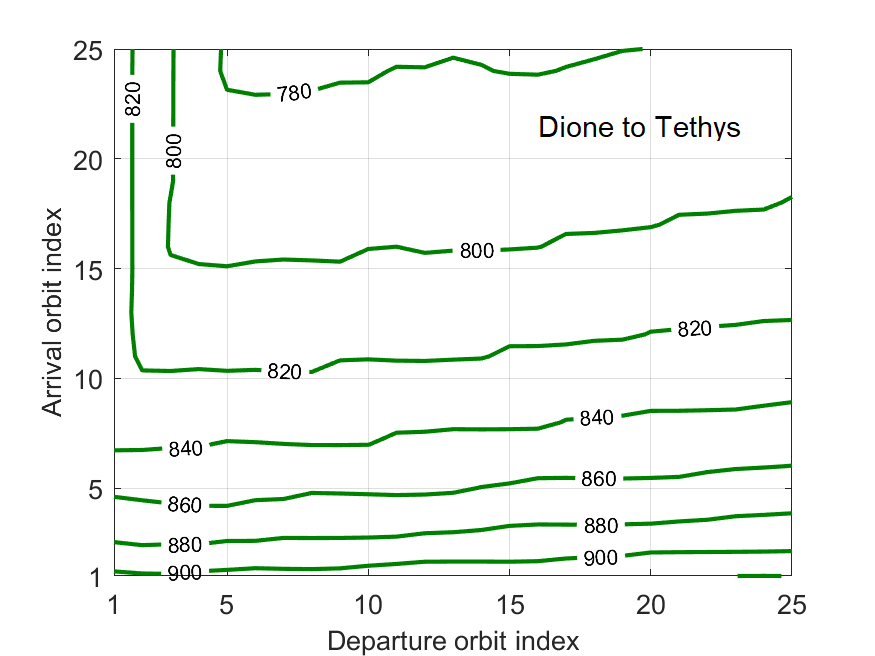} \\
\includegraphics[width=0.62\textwidth]{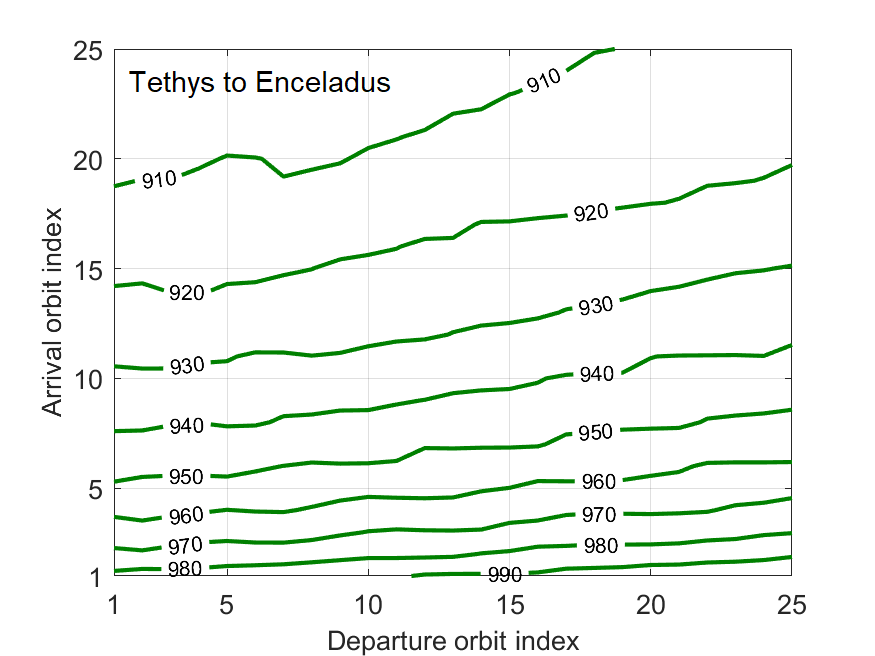} \\
\includegraphics[width=0.62\textwidth]{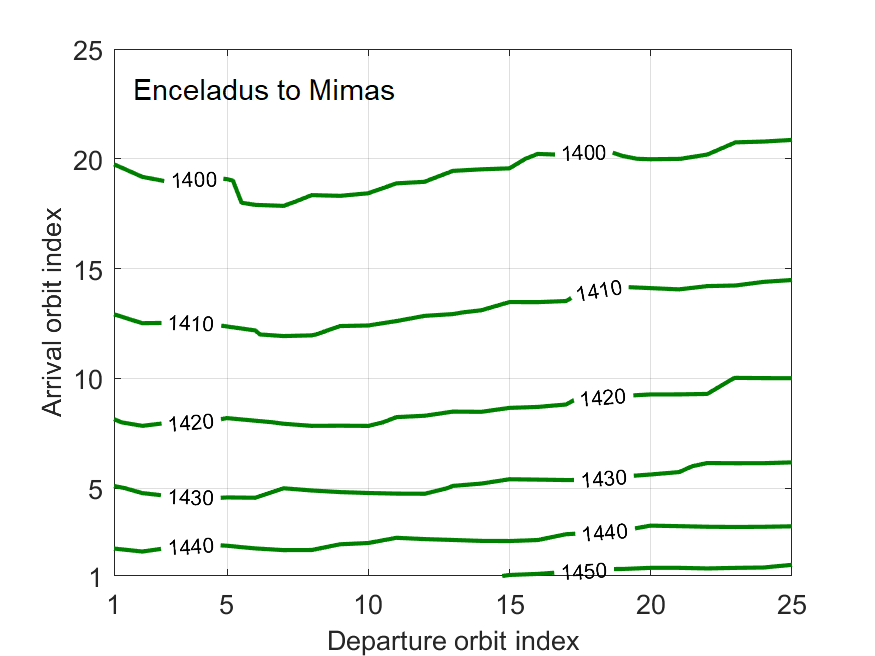} 
\caption{Contour maps of minimum $\Delta V$  (in m/s) transfers for each PLO orbit combination in the three inter-moon segments.}
\label{fig:dV}
\end{figure}

For the inter-moon transfers, $100 \times 25$ sets of initial conditions are available at each end, resulting in $6.25 \cdot 10^6$ candidate trajectories.  
The contour maps of Fig.~\ref{fig:dV} depict the minimum velocity variation ($\Delta V$) over all the trajectories connecting a given pair of PLOs. Due to the large amount of possible combinations of departure and arrival conditions, only the optimal solutions for each orbit pair have been represented (the indices of the optimal arrival and departure PLOs can be found in Table~\ref{tab:inter-moon}). As expected, the lowest cost transfers are between PLOs whose osculating ellipses have reduced energy gaps. However, as stated before, eccentricity is also a factor. The optimal departure trajectories do not have the largest $y$ amplitude (lowest $C_J$, corresponding to index 25). Therefore, they do not minimize the difference of semimajor axes. However, the change in $\Delta V$ between the optimal solution and the transfer between ellipses of minimal energy gap in small (tens of meters per second, at most). Thus, the minimal semimajor axis difference criterion could be used to obtain a preliminary estimate of the total impulse, without having to explore all the trajectory combinations.

Table~\ref{tab:inter-moon} reports the performance characteristics of the optimal solution for each segment of the itinerary (columns 2-4). For the sake of comparison, the LT circle-to-circle transfers between moons (orbital radii $d_i$ and $d_j$) have been computed assuming tangential thrust and Edelbaum's analytical expression for the velocity variation \cite{Edelbaum1961},
\begin{equation}
\Delta V = \displaystyle \left| \sqrt{\frac{GM}{d_i}} - \sqrt{\frac{GM}{d_j}}\right|,
\end{equation}
with $GM$ the gravitational parameter of Saturn. 
The corresponding performance is included in Table~\ref{tab:inter-moon} (columns 5-7) together with that of equivalent Hohmann impulsive maneuvers (columns 8-10, where a specific impulse of 300 s has been assumed).

Most of the transfer time is spent in the space between CoIs. For the sake of completeness, Table~\ref{tab:times_to_CoI} reports the time of flight between the origin/destination PLO and the CoI for the optimal transfers (column 6 with the PLO index given in column 7) as well as the minimum (column 4) and maximum (column 5) time of flight over all the PLOs of the family.

Overall, the tour takes 629 days (1.72 years), consumes 125 kg of propellant and requires a velocity variation of 3.07 km/s. In the three segments, the S/C performs approximately 80, 110 and 230 revolutions around Saturn.  

The advantage of this design over circle-to-circle transfers between moons is noticeable in all respects. The Hohmann maneuvers are obviously very fast, but the associated propellant consumption is more than four times that of the proposed strategy, resulting in insufficient mass budget for the S/C subsystems and science instrumentation. 

\begin{table}[h!]
\caption{Performance comparison (velocity variation $\Delta V$, time of flight $\Delta t$, mass consumption $\Delta m$) among the locally-optimal LT inter-moon transfers between CoIs, the optimal circle-to-circle LT trajectories between moon orbits and equivalent Hohmann impulsive maneuvers.}
\label{tab:inter-moon}
\centering
{\footnotesize\begin{tabular}{lrrrrrrrrr}
\hline
\multicolumn{1}{l}{Segment} & \multicolumn{3}{c}{LT between CoIs} & \multicolumn{3}{c}{LT between moons} & \multicolumn{3}{c}{Hohmann} \\
& $\Delta V$ & $\Delta t\;\;$ & $\Delta m$ & $\Delta V$ & $\Delta t\;\;$ & $\Delta m$ & $\Delta V$ & $\Delta t\;\;$ & $\Delta m$   \\ 
     	    &	 (m/s)        & (day)   & (kg)   &	 (m/s)        & (day)   & (kg) &	 (m/s)        & (day)   & (kg)  \\ \hline
Di-Te & 774 & 170 & 33.8 & 1309 & 283 & 56.1 & 1304 & 1.5 & 251 \\     
Te-En & 901 & 188 & 37.3 & 1285 & 256 & 50.7 & 1281 & 0.8 & 159 \\     
En-Mi & 1392 & 270 & 53.5 & 1656 & 300 & 59.5 & 1650 & 0.6 & 125 \\ 
Total & 3067 & 629 & 125 & 4250 & 839 & 166 & 4235 & 2.9 & 535 \\ \hline   
\end{tabular}}
\end{table} 
\begin{table}[h!]
\caption{Transfer time between origin or destination PLO and the CoI for all the inter-moon transfers: CR3BP (column 1), libration point (column 2),
stability type (column 3, stable for arrivals, unstable for departure segments), minimum (column 4) and maximum (column 5) transfer time over each PLO family, transfer time for the optimal solution (column 6) along with its PLO index (column 7).}
\label{tab:times_to_CoI}
\begin{center}
{\footnotesize\begin{tabular}{llrrrrrrr}
\hline
CR3BP	& L$_i$ & Stability & $(\Delta t_{CoI})_{min}$ & $(\Delta t_{CoI})_{max}$ &  $(\Delta t_{CoI})_{opt}$ & Optimal  	  \\
        &       &           & (day)                &         (day)         &     (day)            & orbit index \\ \hline
SDi     & L$_2$ & Stable    & 1.97                 &  4.69                & 4.19                  & 25 \\  
SDi     & L$_1$ & Unstable  & 1.94                 &  4.67                & 2.18                  & 6 \\ 
STe   & L$_2$ & Stable      &  1.32 & 2.86  & 2.32  & 25 \\ 
STe & L$_1$   &  Unstable   &  1.30 & 2.86  & 1.35 & 2 \\ 
SEn & L$_2$   & Stable & 0.87 & 2.04 & 1.54 & 25 \\ 
SEn & L$_1$ & Unstable & 0.87 & 2.03 & 1.06 & 11 \\ 
SMi & L$_2$ & Stable &  0.57 & 0.90 & 0.83 & 25 \\ \hline 
\end{tabular}}
\end{center}
\end{table} 

\section{Discussion}
\label{sec:disc}
The spacecraft has an initial mass of 1500 kg and carries a Hall effect propulsion system which provides 36 mN of thrust and a specific impulse of 1600 s and is characterized by a power consumption of 640 W. This level of power can be supplied by radioisotope thermoelectric generators, hence the trajectory is not subject to the limitations deriving from the reduced solar radiation flux far from Earth. The interplanetary trajectory optimizes a sequence of unpowered gravity assists at Venus and Earth, along with  propelled and coasting arcs between planets. The objective is to reach Saturn with a hyperbolic excess speed of 1 km/s to unpowered capture around Saturn. The characteristic launch energy is 27.04 km$^2$/s$^2$, 
the total transfer time to Saturn is 12.34 years (from January 21\textsuperscript{st} 2028 to May 24\textsuperscript{th} 2040) and the mass at arrival is 1014 kg. The low approach velocity allows to carry out the orbit insertion at Saturn through a sequence of resonant gravity assists with Titan. The maneuver is unpowered and takes 128 days. It is followed by a powered descent towards Dione, optimized to insert the spacecraft into a planar Lyapunov orbit around the L$_2$ equilibrium point of the Saturn-Dione circular restricted three-body problem, where the tour of the four moons starts. The powered transfer to Dione takes 4.31 years, consuming 312 kg of propellant. This long phase can be exploited to perform scientific observations of the wide region between the orbits of Titan and Dione and make close approaches with Dione and Rhea.
The science orbits around the four target moons are heteroclinic and homoclinic connections between planar periodic orbits around the collinear equilibria L$_1$ and L$_2$ of each Saturn-moon three-body problem. The inter-moon transfers are low-thrust trajectories patching the most favorable conditions among the stable and unstable hyperbolic invariant manifolds of the same periodic orbits used for the exploration of the moons. The science phases require negligible amounts of propellant and can be extended by repeating the same heteroclinic and homoclinic cycles. This possibility derives from the autonomous character of the dynamical model used. The science orbits offer wide surface coverage (up to $\pm 80^{\circ}$ latitude) and long visibility periods  (up to tens of hours) of the targets.  The inter-moon portion of the tour lasts 1.72 years, enabling detailed observations of the E-ring and its environment.

Excluding the science orbits (whose duration can be extended arbitrarily, as mentioned above), the entire mission takes 18.7 years, a duration comparable to that of Cassini/Huygens (20 years) and acceptable for a project of this class.  The proposed concept is novel because it achieves the unprecedented result of inserting the spacecraft into orbit around the four inner moons of Saturn with a propellant mass fraction of 62\%. 
This outcome, which would not be feasible with chemical propulsion technologies, is made possible by the optimal design of low-thrust arcs, the gravitational assistance of intermediate bodies (Venus and Earth in the interplanetary transfer, Titan in the orbit insertion at Saturn) and the dynamical properties of the invariant structures of the circular restricted three-body problem.
The design strategy is holistic, i.e., it deals with the entire transfer taking into account the available onboard resources, the limitations of existing power and propulsion technologies and the desired scientific return of a mission of this importance. The low launch mass makes the proposed concept suitable for small-to-medium-class and even multi-spacecraft missions.

Even if the trajectory is 2D, it is realistic as it enforces phasing constraints derived from the positions of the planets in the interplanetary leg. Moreover, it offers a reasonable degree of fidelity because the dynamical models adopted in every phase take into account the major perturbations. It is expected that only small adjustments will be needed to refine the proposed solution to an $n$-body model.

\section{Conclusions}
\label{sec:conclu}

This study outlined a strategy to explore the four Inner Large Moons of Saturn ---Dione, Tethys, Enceladus and Mimas--- in close proximity using only low-thrust and gravity assist. This is an original concept, impracticable with conventional approaches due to the depth of Saturn's gravity well. The ambitious science goal, inserting the spacecraft into a close orbit around each moon, comes at the expense of increased mission duration. To eliminate impulsive maneuvers completely (which would need a dual propulsion system, adding complexity and weight) a quasi-ballistic capture via a Titan gravity assist is required. To limit the period of the subsequent orbits, the relative velocity upon arrival at Saturn was constrained to 1 km/s. The slow approach velocity to the planet results in a very long interplanetary transfer compared with past missions. Given that the goal of this work is just to demonstrate the feasibility of the concept, this limitation is acceptable.

That being said, there is ample room for improvement. Using the Saturn hyperbolic excess speed as an optimization parameter would establish a trade-off between interplanetary transfer time and the duration of the multi-gravity assists with Titan. This has the potential to decrease the total mission time. Also, shortening the interplanetary phase by extending the time in orbit around to Saturn has some scientific value by itself. Furthermore, for the sake of simplicity, the descent from Titan to Dione has been computed in a sequential fashion. Ballistic trajectories with gravity assists are followed by a propelled phase. It is of course possible to apply thrust between the Titan flybys to shorten the descent. Also, gravity assists with Rhea offer the possibility to accelerate the descent and save propellant.
For simplicity, this study only considered connections between planar Lyapunov orbits of the libration points of the Saturn-moon three-body problems. They have demonstrated great scientific potential as well as substantially reduced cost compared with circle-to-circle inter-moon transfers. Nevertheless, there is also room for improvement in this area by exploring other typologies of libration orbits, such as halo.
Finally, the multi-gravity assist interplanetary phase could also be refined by considering more than three flybys en route to Saturn. This could potentially find sequences of flybys involving Venus, Earth and Jupiter with smaller departure energies and shorter transfer times.

The fact that there is substantial room for improving this results must not be considered a limitation of the concept. In fact, it is one of its greatest strengths. A simplified preliminary analysis has shown that the unprecedented goal of inserting an spacecraft into orbit around the four moons is possible with existing technology. The initial results being easy to improve upon only adds credibility to the concept, and warrants future in-detail analysis.

\section*{Acknowledgements}
The authors are very grateful to Dr. David Morante for his assistance on the use of the global trajectory optimizer and  acknowledge the valuable comments and suggestions of the anonymous reviewer.
The work of E. Fantino, B.~M. Burhani and R. Flores has been supported by Khalifa University of Science and Technology's internal grant CIRA-2021-65 / 8474000413. R. Flores also acknowledges financial support from the Spanish Ministry of Economy and Competitiveness ``Severo Ochoa Programme for Centres of Excellence in R\&D'' (CEX2018-000797-S). In addition, E. Fantino received partial support from the Spanish Ministry of Science and Innovation under projects PID2020-112576GB-C21 and PID2021-123968NB-I00. M. Sanjurjo-Rivo acknowledges fund PID2020-112576GB-C22 of the Spanish Ministry of Science and Innovation.

\bibliography{Saturn_Biblio_R1}

\end{document}